\documentclass[twocolumn,prb,superscriptaddress,floats,nobibnotes,notitlepage,citeautoscript]{revtex4-1}

\usepackage[pdftex]{graphicx,hyperref}
\usepackage{amsmath,bm}
\usepackage{bbold}
\usepackage{color}
\usepackage{wasysym}
\usepackage{amssymb}
\usepackage{multirow}
\usepackage{float}
\usepackage{booktabs}
\usepackage{ulem}
\usepackage{dsfont}

\def\<{\langle}
\def\>{\rangle}

\newcommand{\ve}[1]{\boldsymbol{#1}}

\definecolor{DarkRed}{RGB}{193,40,40}

\begin{document}

\title{Bandwidth controlled quantum phase  transition    between  an  easy-plane  quantum spin Hall state  and  an  s-wave  superconductor}

\author{Disha Hou}
\affiliation{\mbox{Department of Physics, Beijing Normal University, Beijing 100875, China}}
\author{\firstname{Yuhai} \surname{Liu}}
\affiliation{\mbox{Beijing Computational Science Research Center, 10 East Xibeiwang Road, Beijing 100193, China}}
\affiliation{\mbox{School of Science, Beijing University of Posts and Telecommunications, Beijing 100876, China}}
\author{\firstname{Toshihiro} \surname{Sato}}
\affiliation{\mbox{Institut f\"ur Theoretische Physik und Astrophysik, Universit\"at W\"urzburg, 97074 W\"urzburg, Germany}}
\author{\firstname{Wenan} \surname{Guo}}
\email{waguo@bnu.edu.cn}
\affiliation{\mbox{Department of Physics, Beijing Normal University, Beijing 100875, China}}
\affiliation{\mbox{Beijing Computational Science Research Center, 10 East Xibeiwang Road, Beijing 100193, China}}
\author{Fakher F. Assaad}
\email{assaad@physik.uni-wuerzburg.de}
\affiliation{\mbox{Institut f\"ur Theoretische Physik und Astrophysik, Universit\"at W\"urzburg, 97074 W\"urzburg, Germany}}
\affiliation{\mbox{W\"urzburg-Dresden Cluster of Excellence ct.qmat, Am Hubland, 97074 W\"urzburg, Germany}}
\author{Zhenjiu Wang }
\email{zhwang@pks.mpg.de}
\affiliation{Max-Planck-Institut f\"ur Physik komplexer Systeme, Dresden 01187, Germany}

\begin{abstract}

The  quantum spin Hall  state  can  be understood  in terms of  spontaneous  O(3)  symmetry breaking. Topological skyrmion configurations of the O(3) order parameter vector carry  a charge 2e,  and as shown  previously,  when they  condense,  a superconducting state is generated.   
We show that this topological route to superconductivity survives easy-plane anisotropy. Upon reducing the O(3) symmetry to O(2)$\times$ Z$_2$, 
skyrmions give  way to   merons that  carry  a unit charge. On  the  basis of  large-scale   auxiliary  field  quantum Monte  Carlo   simulations,   we  show    that  at the
particle-hole   symmetric  point,   we  can   trigger  a continuous and direct transition between the  quantum spin Hall state and  s-wave  
superconductor by condensing  pairs of  merons. This statement is valid in both strong and weak  anisotropy limits.   
Our results can be interpreted in  terms of   an easy-plane  deconfined quantum critical point.  
However, in contrast to the previous studies in quantum spin models,  our 
realization of this  quantum critical point conserves $U(1)$  charge, such  that  skyrmions  are   conserved.   

\end{abstract}

\maketitle

\section{Introduction}
Topology  is  a   key  factor  for  understanding  phase  transitions.   In the  Kosterlitz-Thouless   theory  \cite{Kosterlitz73},  an   O(2)  local  order  parameter  in  two-dimensional space  allows  for  the  definition of the vortex,  the   proliferation  of    which   drives  the   transition. 
Let us stay  in   two-dimensional space,  $\ve{x} = (x,y)$,  but  consider  an  O(3)  local  order  parameter   $\ve{n}(\ve{x})$  with  the unit norm.  
 This  combination   of  space    and  order  parameter  defines a  winding   number,~\cite{fradkin_book_2013} 
\begin{equation}
\label{top_index.eq}
      \frac{1}{4\pi}  \int  d^{2}\ve{x}  \,  \ve{n}   \cdot   \partial_x   \ve{n}   \times   \partial_y \ve{n}.
\end{equation}
For  smooth   configurations,  
this  quantity is  quantized  and  counts    the   winding of the unit   vector  on the   unit   sphere:   a  skyrmion.  
We  can  now   reduce  the  O(3)  symmetry  to   O(2) $\times$  Z$_2$.     In the context  of  spin  systems,  this  would  correspond  to  restricting the  
O(3) symmetry to O(2) transformations around,  say,  the  z-axis  and  a change  of  sign  of the third  component of the   $\ve{n}$-vector.   
Assuming that energetics favor the  $\ve{n}$-vector   to  be in-plane (i.e., vanishing z-component)  then  the   topological   excitations 
will   correspond  to  vortices  in the  x-y  plane.    Due  to the normalization condition,     the $\ve{n}$-vector  at the  core of the vortex   
will have to point   along the z-direction.  
Since  the  $\ve{n}$-vector lies in the x-y plane  at infinity,  the integrand  of  Eq.~\ref{top_index.eq}   vanishes  at  infinity   and  the  
integral   takes  half-integer  values:  a  meron. 

The   above  considerations  acquire  different  interpretations  depending upon the  specifics of the   local  order  parameter.   In this paper, it 
 corresponds to  the \textit{ order-parameter}  of the  quantum-spin Hall  state.    In  particular,  let  
$  \hat{H}_0   =    -v_F \sum_{\ve{p},\sigma, i=1,2}    \hat{\ve{\Psi}}^{\dagger}_{\sigma} (\ve{p})  i  p_i \gamma_0  \gamma_i   \hat{\ve{\Psi}}^{}_{\sigma}(\ve{p}) $   
be  the  Hamiltonian   akin  to 
graphene in the absence of interactions \cite{Neto_rev}, using  the  notation of  Ref.~\onlinecite{Herbut09}.    Inclusion of the  quantum  spin Hall mass term  leads  to: 
\begin{equation}
	   \hat{H}    =  \hat{H}_0  +  \int_V d \ve{x} \ve{N}(\ve{x})   \cdot  \hat{ \ve{\Psi}}^{\dagger}_{\sigma}(\ve{x})  i  \gamma_0 \gamma_3 \gamma_5  \ve{\sigma}_{\sigma,\sigma'} \hat{\ve{\Psi}}^{}_{\sigma'}(\ve{x}). 
\end{equation}
The  order parameter  $\ve{N}$    can be normalized  to unity  $\ve{n}   = \ve{N}/|\ve{N}| $ if,    as will be  the  case   in  our model,  the  single-particle gap, that is proportional to the length of $\ve{N}$,  does not  vanish.   
 Furthermore,  $\ve{n}$  is odd  under  charge  conjugation    such that   the expression in 
Eq.~(\ref{top_index.eq})   carries the  same  quantum numbers as  the  charge  density $\rho$  measured  with  respect to  half-filling. In  particular, in  Ref.~\onlinecite{Grover08}   it is shown that  
\begin{equation} 
    \rho(\ve{x})   = \frac{2 e}{4\pi}  \ve{n}   \cdot   \partial_x   \ve{n}   \times   \partial_y \ve{n}
\end{equation}
such  that  the  skyrmion (meron pairs)  carry  charge  2e. Since  topological  defects of one phase carry  the charge of the other,  their  
proliferation will  lead  to a symmetry-broken state.   In the  above discussion,  we  have  considered    the quantum spin Hall (QSH)  state  
where the skyrmions (or  pairs of  merons)   carry  charge  2e   and  their  proliferation leads to  a  superconducting state.  Alternatively,   
one  could  consider   the   three   spin-density  wave  mass terms.  
In this case, the skyrmion creation/annihilation process would acquire a phase under the spatial rotation of the lattice.
This  proliferation of  skyrmions   is  the  essence  of   so-called `deconfined' quantum   critical points   (DQCPs)~\cite{Senthil04_1, Senthil04_2, WangC17}.  

The   above  discussion  takes place in the continuum,  and  some  type of  regularization  is needed to carry out  numerical  simulations.  
Starting  with  magnetic 
phases, where the  topological  defects carry a  U(1)  rotational charge,   a lattice  regularization   leads to additional symmetry breaking   
 terms that may lead to subtleties since they do not  exist in the  expected IR field theory~\cite{Sandvik07, Shao15}.   
  For  instance  for a  square lattice  regularization scheme,   U(1)    rotational  symmetry  gives way  to a  $ \mathbb{Z}_4$  invariance of  the  valence bond  solid (VBS)  state.  This   regularization-induced  symmetry  reduction  introduces  novel  operators  that have
  to be argued to be irrelevant  at  the  critical point. 
  In particular, in the  framework of  the CP$^{1}$ field  theory of  spinon  coupled  to a U(1)  gauge  field~\cite{Senthil04_1},  this  symmetry  reduction allows 
  for  the  creation of  quadruple  monopoles  of  the gauge  field. 
  
  Although the $Z_4$ lattice symmetry-breaking field is relevant  in the VBS state,      
a necessary condition for the continuous nature  of the  transition into an xy-antiferromagnetic (AFM) phase  
is that this symmetry-breaking field is irrelevant at   the critical point.  
Numerically, AFM-VBS phase transitions in the easy-plane case   
have clear first-order signatures in most cases~\cite{KUKLOV_2006, Emidio_2016, Sandvik_2002_Stripe, Kragset_2006, Sen_2007, Emidio_2017, Desai_2020}.        
Among numerical works, Desai et. al.~\cite{Desai_2020}, in p  
emphasize  the absence of the continuous transition in any easy-plane spin system: the authors 
generally   
claimed the absence of the easy-plane deconfined fixed point without considering the effect of 
quadruple monopoles.  
However, could it be that the $Z_4$ symmetry-breaking field introduces a runaway flow, leading 
to a first-order transition?

Instead  of  encoding  the U(1)  symmetry  as a  rotational invariance,  one  can encode it in  terms of charge  conservation.  
Importantly,  charge  conservation will not be broken by  lattice  regularization.
Following the work of Ref.~\onlinecite{Liu18, Zwang_doping}  we set up a set of designer Hamiltonians  which have the 
 potential to realize
an easy-plane DQCP without quadruple monopoles.
A dynamically generated QSH insulating state which spontaneously breaks the O(2) spin rotational
symmetry  emerges from a Dirac semi-metal via a spin-orbital interaction.
Our particular interest lines in the phase transition between the  QSH and  s-wave superconducting (SSC) states.

The fermion basis introduces a simple but much more 
provoking picture: meron defects of the QSH order parameter which carry a unit of electron charge are the fundamental
excitation at the critical point; on the other side of the transition, the condensation of meron-pair creation/annihilation operators forms the superconducting state.  
Importantly, the U(1) charge conservation broken by the SSC phase is an exact symmetry of our lattice Hamiltonian, meaning that  
quadruple monopoles are absent by definition. 

The aim of this work is to systematically search  for the existence of an easy-plane DQCP without monopoles.
The continuity of phase transitions  does not only depend on symmetry.~\cite{Domany1984,Guo_2002}
In our lattice model, a  model parameter that is related to the \textit{strength} of the easy-plane anisotropy continuously
tunes the energy gap of meron configurations of a QSH order parameter.
Regardless  of  the strength of the anisotropy  and  in contrast  to  lattice spin models,
our model shows no obvious signs of first-order transitions.
We argue that this transition flows to the easy-plane DQCP.

The paper is organized as follows. In Sec.~\ref{Sec:Model} we introduce our lattice Hamiltonian, the quantum Monte Carlo algorithm, as well as basic observables.
The numerical results are shown in Sec.~\ref{Sec:Results}, beginning with the ground state phase diagram and followed by 
a  detailed investigation of  the nature of  the phase transitions.
Finally, we draw conclusions and give an outlook in Sec.~\ref{Sec:Conclusions}.

\section{Model and Methods}
\label{Sec:Model}

We consider a model of Dirac fermions in $2+1$
dimensions on the honeycomb lattice with Hamiltonian
\begin{equation}\label{Eq:Ham_T}
\begin{aligned}
 \hat{H}_t   = - t  \sum_{ \langle \bm{i}, \bm {j} \rangle } (\hat{\ve{c}}^{\dagger}_{\bm{i} } \hat{\ve{c}}^{\phantom\dagger}_{\bm{j}}  + H.c.).
\end{aligned}
\end{equation}
Here, the spinor
$\hat{\boldsymbol{c}}^{\dag}_{\ve{i}} =
\big(\hat{c}^{\dag}_{\ve{i},\uparrow},\hat{c}^{\dag}_{\ve{i},\downarrow}
\big)$ where $\hat{c}^{\dag}_{\ve{i},\sigma} $ creates an electron  in a  Wannier state  centered around lattice  
site $\ve{i}$ with $z$-component of spin $\sigma$. This  term
accounts for nearest-neighbor hopping.   The interaction term that we consider reads:
\begin{equation}\label{Eq:Ham_V}
\begin{aligned}
 \hat{H}_{\lambda}  = & -\lambda \sum_{\varhexagon}
   \left[  \left( \sum_{\langle \langle \bm{i} \bm{j} \rangle \rangle  \in \varhexagon }   \hat{J}^x_{\bm{i},\bm{j}} \right)^2
   +    \left( \sum_{\langle \langle \bm{i} \bm{j} \rangle \rangle  \in \varhexagon }   \hat{J}^y_{\bm{i},\bm{j}} \right)^2     \right.      \\
 &  \phantom{=\;\;} \left.
 +  \Delta  \left( \sum_{\langle \langle \bm{i} \bm{j} \rangle \rangle  \in \varhexagon }   \hat{J}^z_{\bm{i},\bm{j}} \right)^2     \right]
\end{aligned}
\end{equation}
where
$ \hat{ \bm{J} }_{\bm{i},\bm{j}} \equiv i \nu_{ \bm{i} \bm{j} }
\hat{\ve{c}}^{\dagger}_{\bm{i}} \bm{\sigma}
\hat{\ve{c}}^{\phantom\dagger}_{\bm{j}} + H.c.$ The components of $\boldsymbol{\sigma}=(\sigma^x,\sigma^y,\sigma^z)$ are the Pauli
spin-1/2 matrices. This term is a plaquette
interaction involving next-nearest-neighbor pairs $\langle \langle \bm{i} \bm{j} \rangle \rangle$ of sites and phase factors
$\nu_{\boldsymbol{ij}}=\pm1$ identical to the Kane-Mele model
\cite{KaneMele05b}, see also Ref.~\onlinecite{Liu18}.

The Hamiltonian $\hat{H}=\hat{H}_t +\hat{H}_\lambda$ with $\Delta=1$ has been studied in Ref.~\onlinecite{Liu18}.
A dynamically generated QSH insulator that breaks $SU(2)$ spin rotational symmetry spontaneously is found at
intermediate  interacting strength ($\lambda$), separating a Dirac
semi-metal(DSM) state at small $\lambda$ and an SSC state at large
$\lambda$. The DSM-QSH transition belongs to the
Gross-Neveu Heisenberg universality class \cite{Gross74}  whereas
the QSH-SSC transition falls into the class of DQCP.
 In  the current work, we focus on the case of $ \Delta \in [ 0, 1 ) $
where the $SU(2)$ spin rotational symmetry is reduced to $U(1) \times Z_2$.

We used the ALF (Algorithms for Lattice Fermions)
implementation~\cite{alfcollaboration2021alf} of the well-established auxiliary-field quantum
Monte Carlo (QMC) method~\cite{Blankenbecler81,White89,Assaad08_rev}.  Because
$\lambda > 0$ and $\Delta>0$, we can use a real Hubbard-Stratonovich decomposition for the
perfect square term. We set the imaginary time interval $\Delta \tau=0.2$ and choose a symmetric Trotter decomposition 
to ensure the hermiticity of  the imaginary time propagation in the  Monte Carlo simulations.\cite{Liu18}
Additionally, a checkerboard decomposition is applied to the exponential of hopping matrix $\hat{H}_t$.
For each field configuration, time-reversal symmetry  and   charge are conserved. Hence the eigenvalues of the  fermion  determinant 
occur  in  complex conjugate pairs,  and we  do not suffer from the negative sign problem \cite{Wu04}. 
We simulated lattices with $L\times L $ unit cells (each containing two Dirac fermions) and periodic boundary conditions.
Following our previous work~\cite{Zwang_doping},  we used a projective version of the algorithm (PQMC)
\cite{Sugiyama86,Sorella89,Assaad08_rev}.
This algorithm is based on the  form: 
\begin{equation}
	\langle \hat{O} \rangle   =  \lim_{\theta \rightarrow \infty} \frac{\left< \Psi_T|  e^{-\theta \hat{H}}  \hat{O}  e^{- \theta \hat{H}} |\Psi_T \right>}
	{\left< \Psi_T|  e^{-2\theta \hat{H}}  |\Psi_T \right>}.
\end{equation}
Provided  that  the  trial  wave function $\left. |\Psi_T \right>$  is not orthogonal to the ground state,  $\langle \hat{O} \rangle   $  corresponds  to the 
ground state   expectation value of the  observable $\hat{O}$. 
 To avoid the negative sign problem, we consider the same time-reversal symmetric trial wave function as the one used in
Ref.~\onlinecite{Zwang_doping}.
We explicitly checked the projection convergence to  the  ground state at each  system size and each $\Delta$:
simulations are performed at $L=6, 9, 12, 15, 18$ and the values of projection length $\theta$ for ground state calculation are listed in Tab.~\ref{tab:theta}.
\begin{table}[tp]
\caption{\label{tab:theta}
    Projection length $\theta$ at different values of $\Delta$ and $L$.
}
  \centering
  \begin{tabular}{c c c c}
    \toprule
     \hline
    \multirow{2}{*}{lattice size}&
    \multicolumn{3}{c}{anisotropy strength}\cr
    &  ~~~$ \Delta=0.1 $~~~~&~~~~$\Delta=0.5$~~~~&~~~~$ \Delta=0.75$  ~~~~~\cr
    \midrule
    \hline
    $L=6$ &15&15&15\cr
    $L=9$ &21&21&21\cr
    $L=12$&24&24&24\cr
    $L=15$&42&36&36\cr
    $L=18$&42&36&36\cr
     \hline
    \bottomrule
    \end{tabular}
\end{table}

The basic measurements in our QMC simulations are  equal time correlation functions in real space: 
\begin{equation}
\begin{aligned}\label{Eq:def_S_r}
   S^O( \bm{r}  )
   \equiv  \frac{1}{L^2}  
  \sum_{n} \sum_{\boldsymbol{r}'}  
   \langle \hat{O}^\dagger_{\bm{r}, n }  \hat{O}^{\phantom\dagger}_{\bm{r} + \bm{r}', n }   \rangle ,
\end{aligned}
\end{equation}
and the 
structure factor:
\begin{equation}
\begin{aligned}
   S^O_{m, n} ( \bm{q} )
   \equiv \frac{1}{L^2} \sum_{\boldsymbol{r},\boldsymbol{r'}} e^{\mathrm{i}\boldsymbol{q}\cdot(\boldsymbol{r}-\boldsymbol{r}')}
   \langle \hat{O}^\dagger_{\bm{r},m }  \hat{O}^{\phantom\dagger}_{\bm{r}', n }   \rangle ,
\end{aligned}
\end{equation}
where $\hat{O}_{\bm{r},n}$ is a local operator   with $\bm{r}$ denoting the unit cell  and  $n$   denoting the intra unit-cell    dependence that  we  will  
refer to  as orbital.

\begin{figure}
\centering
\includegraphics[width=0.5\textwidth]{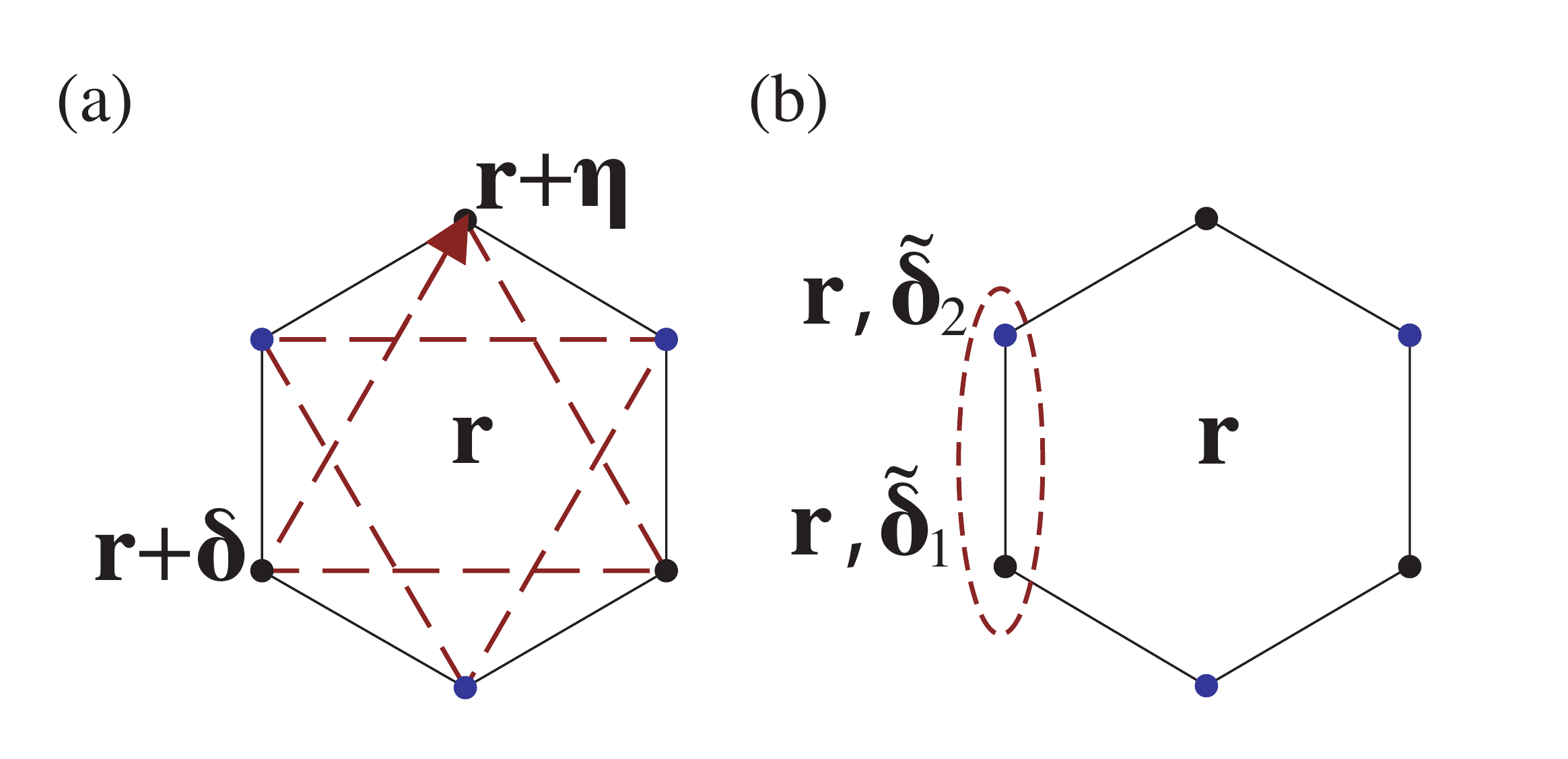}
\caption{\label{fig:operator}
Honeycomb plaquette illustrating (a) the spin-orbit coupling operator $\hat{ \bm{J} }_{\ve{r} +
  \ve{\delta}_n,\ve{r} + \ve{\eta}_n}$ and (b) the pairing operator $\hat{\eta}^{+}_{\ve{r},\ve{\tilde{\delta}_a}}$.
}
\end{figure}

For  instance,  the spin-orbit coupling operators
correspond  to  $\hat{\bm{O}}_{ \ve{r}, n }  =  \hat{ \bm{J} }_{\ve{r} + \ve{\delta}_n,\ve{r} + \ve{\eta}_n}$. 
Here $n$ runs over the six next-nearest neighbor bonds of the corresponding
hexagon with legs $\ve{r} + \ve{\delta}_n$ and $\ve{r} + \ve{\eta}_n$ ($n=1,2,...6$), see Fig. \ref{fig:operator}.  To detect QSH ordering which breaks the $ O(2)$ spin rotational symmetry, we calculate the 
structure factor matrix associated with the $X$ and $Y$ components of $\hat{\bm{O}}_{\ve{r},n}$
\begin{equation}\label{Eq:sf_qsh}
\begin{aligned}
   S^{\text{QSH} }_{m, n} (\boldsymbol{q})
        \equiv   \frac{1}{L^2} \sum_{\boldsymbol{r},\boldsymbol{r'}}    e^{\mathrm{i}\boldsymbol{q}\cdot(\boldsymbol{r}-\boldsymbol{r}')}  
        \langle  \hat{ O }^X_{\boldsymbol{r}, m} \hat{  O  }^X_{\boldsymbol{r'}, n }
    +
     \hat{ O }^Y_{\boldsymbol{r},  m}  \hat{ O }^Y_{\boldsymbol{r'},  n}
    \rangle,
\end{aligned}
\end{equation}
with $m,n=1,2,...6$.
We also consider the structure factor matrix associated with the $Z$ component of $\hat{\bm{O}}_{\ve{r},n}$
\begin{equation}
\begin{aligned}\label{Eq:def_qshZ}
   S^{\text{QSH}_Z }_{m, n} (\boldsymbol{q})
        \equiv   \frac{1}{L^2} \sum_{\boldsymbol{r},\boldsymbol{r'}}    e^{\mathrm{i}\boldsymbol{q}\cdot(\boldsymbol{r}-\boldsymbol{r}')}
        \langle  \hat{ O }^Z_{\boldsymbol{r}, m}  \hat{  O  }^Z_{\boldsymbol{r'}, n }
    \rangle.
\end{aligned}
\end{equation}
The physical meaning of this quantity will be discussed in Sec.~\ref{Sec:Results}.

To detect SSC ordering which breaks $U(1)$ charge conservation, we consider the following structure factor matrix:
\begin{equation}\label{Eq:sf_ssc}
\begin{aligned}
   S^{\text{SSC} }_{a, b} (\boldsymbol{q})
        \equiv    \frac{1}{L^2} \sum_{\boldsymbol{r},\boldsymbol{r'}}     e^{\mathrm{i}\boldsymbol{q}\cdot(\boldsymbol{r}-\boldsymbol{r}')}      
	 [ \langle  \hat{ \eta }^{+}_{\boldsymbol{r},\boldsymbol{ \widetilde{\delta} }_a } \hat{ \eta  }^{-}_{\boldsymbol{r'},\boldsymbol{ \widetilde{ \delta} }_b }  \rangle +
	 \langle  \hat{ \eta }^{- }_{\boldsymbol{r},\boldsymbol{ \widetilde{\delta} }_a }  \hat{ \eta  }^{+}_{\boldsymbol{r'},\boldsymbol{ \widetilde{\delta} }_b }
	\rangle],
\end{aligned}
\end{equation}
with $a,b=1,2$, denoting the A(B) sublattice, where the s-wave pairing operator is defined as
\begin{equation}
    \hat{\eta}^{+}_{\ve{r},\ve{\tilde{\delta}_a}} = \hat{c}^{\dagger}_{\ve{r}
  +\ve{\tilde{\delta}_a},\uparrow} \hat{c}^{\dagger}_{\ve{r}
 +\ve{\tilde{\delta}_a},\downarrow}.
\end{equation}
Here $\tilde{\delta}_a$ runs over the two orbitals in unit cell $\ve{r}$,
see Fig. \ref{fig:operator}.

We use Eq.~\ref{Eq:sf_qsh} and Eq.\ref{Eq:sf_ssc} to calculate the order parameter:

\begin{equation}
    m^O=\sqrt{\frac{\Lambda_1(S^O(\bm{Q}))}{L^2}}
\end{equation}

Here, $\Lambda_1()$ indicates the largest eigenvalue of the corresponding matrix in orbital space, $O$ denotes QSH and SSC order parameters, and
 $\bm{Q}=(0,0)$. 
The   corresponding  eigenvector  will  determine  the orbital  structure.  This is of  particular importance  for the QSH  state  since    we  expect it 
to  reflect  the sign structure  $\nu_{\ve{i}\ve{j}} $   of  the Kane-Mele model.

To locate the critical points and study the critical properties,
after diagonalizing the corresponding structure factors,
 we calculated the renormalization-group invariant correlation ratio
\begin{equation}
    R^{O}  \equiv  1 - \frac{ S^O ( \bm{Q} + \Delta \bm{q} )  }{  S^O ( \bm{ Q } ) }
\end{equation}
using the largest eigenvalue  $S^{O}$ ($O$ = QSH, QSH$_z$, SSC); $\bm{Q} = (0, 0)$ is the ordering wave vector and  $\bm{Q} + \Delta \bm{q}$ is a neighboring wave vector with  $ | \Delta \bm{q}  | = \frac{4 \pi}{ \sqrt{3} L} $. 
By definition, $R^O\to 1$ for $L\to\infty$ in the corresponding ordered state, whereas $R^O\to 0$ in the disordered state.  
At the critical point, $R^O$ is scale-invariant for sufficiently large $L$ so that the results for different system sizes cross.

\section{Quantum Monte Carlo results}
\label{Sec:Results}

In this section we will first provide the ground state phase diagram and then will proceed to investigate the nature of the phase transitions.

\subsection{Ground state phase diagram}
\label{Sec:Phase_diagram}

As mentioned  previously, we are interested in the parameter range of $\Delta \in [0, 1 )$  where 
the spin rotational symmetry of the Hamiltonian $\hat{H}=\hat{H}_t +\hat{H}_\lambda$ is lowered to $U(1) \times Z_2$.
The DSM and SSC states found in the $SU(2)$ symmetric case~\cite{Liu18} ( $\Delta =1$ ) 
 are naturally 
 stable against weak easy-plane anisotropy $ \Delta \approx 1 $ 
 since  both states are  spin rotational invariant.   Furthermore,   since   time  reversal  symmetry  
 is  not broken by   our  symmetry  reduction,  we  expect the  QSH   phase  to be  equally stable. 
To  confirm the  above,  we  can use the  mean-field  approach  introduced  in Ref.~\onlinecite{Zwang_doping}  
that  carries over  to  the  anisotropic case.  
Due to the Dirac nature of the kinetic term in  
Eq.~\ref{Eq:Ham_T}, we foresee the robustness of the DSM phase in the weakly interacting case.  
On the  other hand, the \textit{attractive} nature of $\hat{H}_{\lambda}$ term ( for $\Delta \geq 0$, $\lambda> 0$ ) suggests that the mean-field picture in the large $\lambda$ case~\cite{Zwang_doping} will still favor an SSC instability.  
Finally, the dynamically generated QSH state at intermediate values of $\lambda$  
will be  restricted to the $U(1)$ plane in the current case.  
We present the mean-field phase diagram in Fig.~\ref{fig:MF_Diagram}.   The  details 
of  the   calculations  are  summarized in Appendix~\ref{Appendix}. 
It is worth  mentioning that, in   the  mean-field analysis,   
the  QSH and SSC orderings coexist in a large $\lambda$ region
of  the phase  diagram,  see  Fig.~\ref{fig:MF_Diagram}. 
This is a natural consequence  of the anti-commuting nature of the two Dirac masses.     
In this  case, the fermion band gap  is   given  by  the norm of the  four-component     
order  parameter    accounting for the  QSH ($m_{\text{QSH}}$) and  SSC  ($m_{\text{SSC}}$) orders:
\begin{equation}
    \Delta_{\text{MF}} \propto \sqrt{
    m^2_{\text{QSH}} + m^2_{\text{ SSC}} }.
\end{equation} 
Hence,  developing  superconducting  ordering  
in the background of a QSH state can simply minimize the mean-field free energy~\cite{Zwang_doping} 
in  the \textit{ large } $\lambda$  limit.

\begin{figure}[H]
\centering
\includegraphics[width=0.48\textwidth]{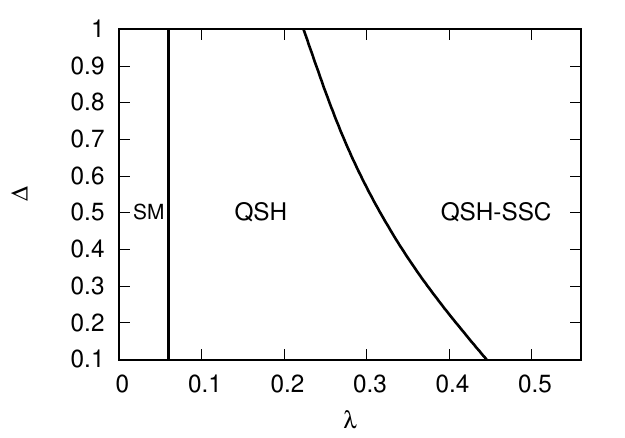}
\caption{\label{fig:MF_Diagram}
Mean-field phase diagram at zero temperature as a function of $\Delta$ and $\lambda$.
}
\end{figure}

\begin{figure}[h]
\centering
\includegraphics[width=0.48\textwidth]{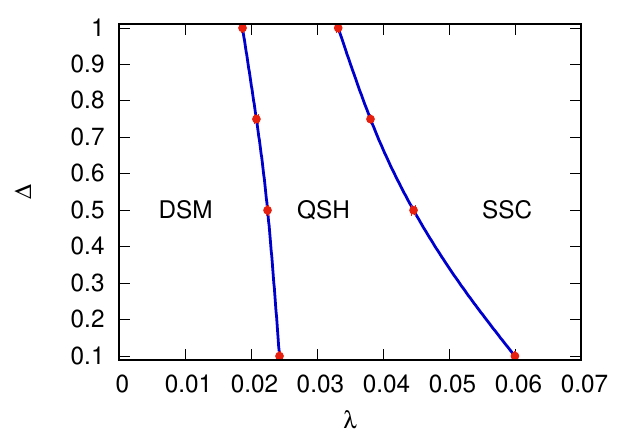}
\caption{\label{fig:Phase_diagram}
Ground state phase diagram in the $(\Delta, \lambda)$ plane which covers the range of $\Delta \in [0.1, 1]$.  The DSM-QSH and QSH-SSC phase boundaries at $\Delta=0.1, 0.5$,
and $0.75$ are estimated from PQMC simulations in this paper.  The phase boundaries of $SU(2)$ symmetric Hamiltonian  (the overline of $\Delta=1$ ) are from Ref~\cite{Liu18}. 
}
\end{figure}

We summarize the exact ground state phase diagram in  $ (\Delta, \lambda ) $ plane based on QMC results in Fig.~\ref{fig:Phase_diagram},
where the overline $\Delta=1$ corresponds to the $SU(2)$ symmetric model that
was studied in Ref.\onlinecite{Liu18}. 
Generally speaking, we found a DSM state at weak interaction (small $\lambda$ region), an SSC state at strong interaction (large $ \lambda $ region),  as well as a $U(1)$ broken QSH state at an intermediate region.

\begin{figure}[htbp]
\centering
\includegraphics[width=0.45\textwidth]{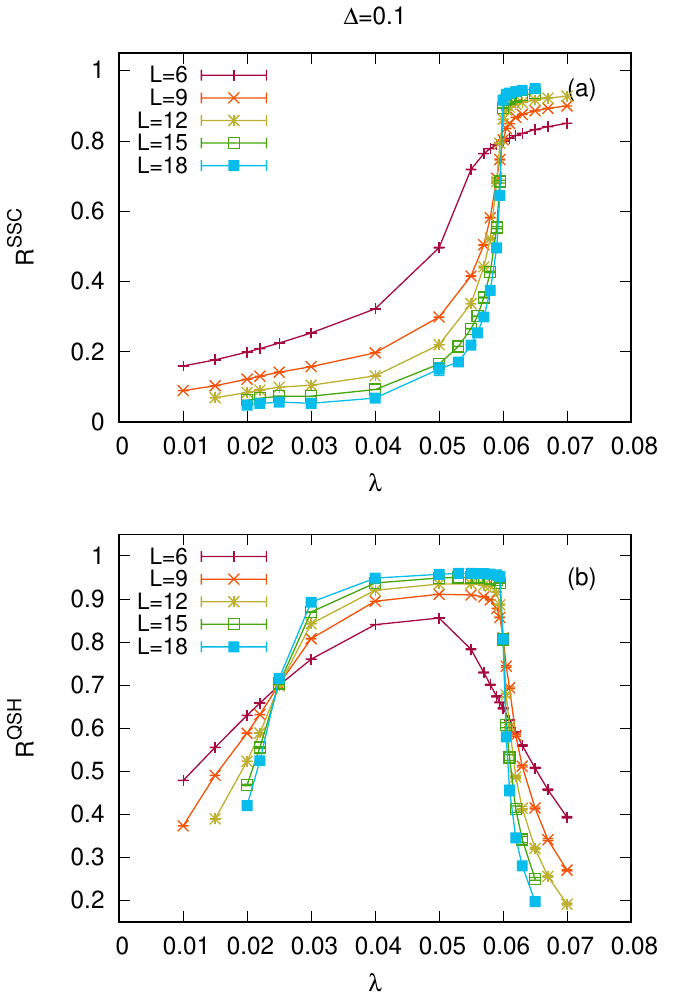}
\caption{\label{fig:Ratioeq_0.1}
Equal-time correlation ratio $R^{\rm SSC}$ (a) and $R^{\rm QSH}$ (b) as a function of $\lambda$ for $\Delta=0.1$.
}
\end{figure}

\begin{figure}[htbp]
\centering
\includegraphics[width=0.45\textwidth]{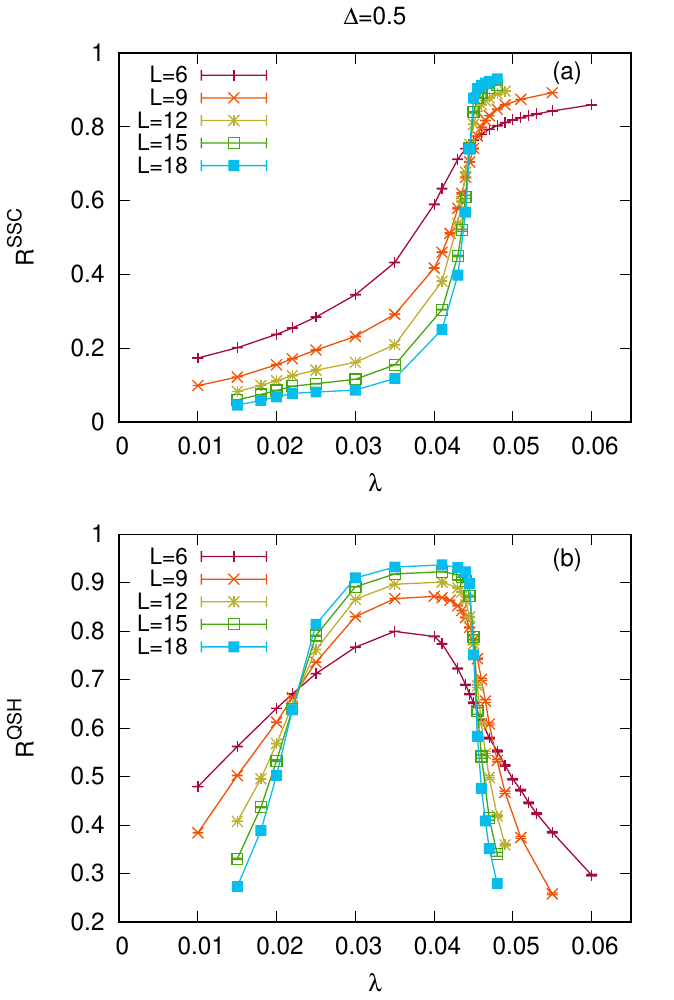}
\caption{\label{fig:Ratioeq_0.5}
Same as Fig.~\ref{fig:Ratioeq_0.1} for $\Delta=0.5$.
}
\end{figure}

\begin{figure}[htbp]
\centering
\includegraphics[width=0.45\textwidth]{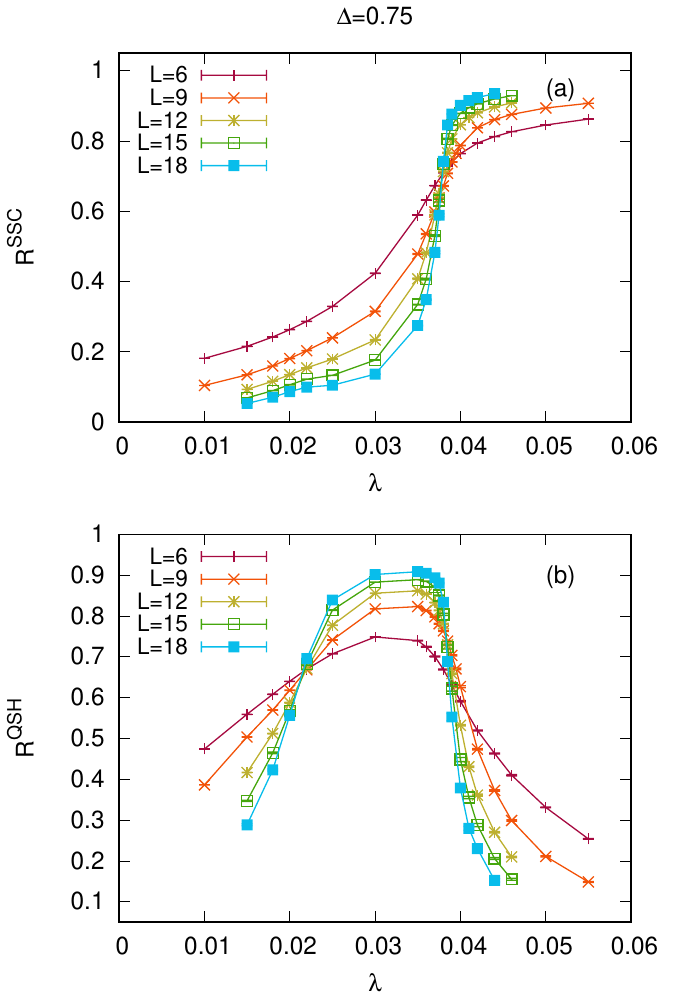}
\caption{\label{fig:Ratioeq_0.75}
Same as Fig.~\ref{fig:Ratioeq_0.1} for $\Delta=0.75$.
}
\end{figure}

\begin{figure}[ht]
    \includegraphics[width=0.48\textwidth]{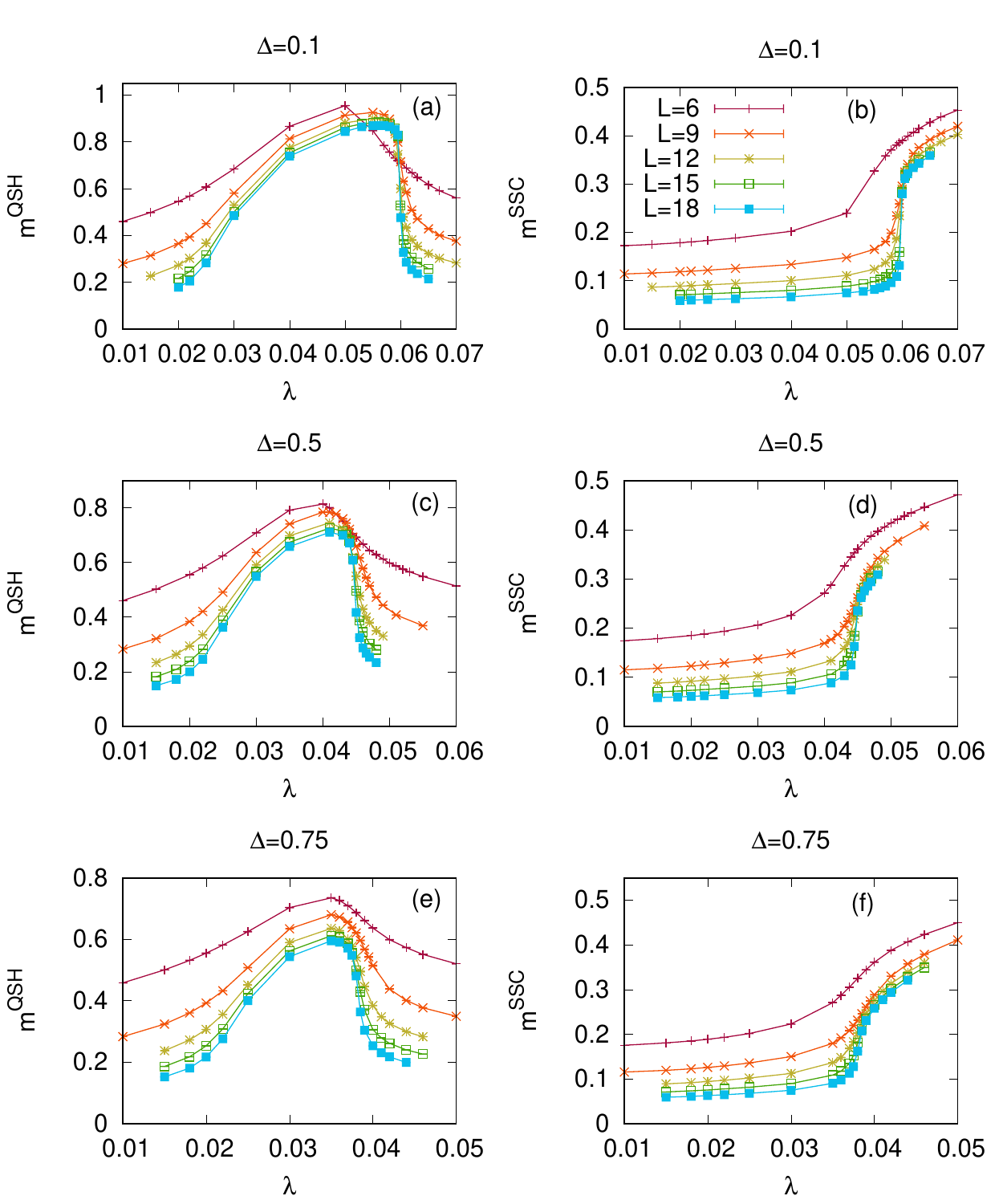}
    \caption{QSH and SSC order parameter as a function of $\lambda$, for 
    $\Delta=0.1$ ((a) and (b)), $0.5$ ((c) and (d)) and $0.75$ ((e) and (f)). }
    \label{fig:m}
\end{figure}

The numerical simulations that we performed cover three  horizontal lines as a function of $\lambda$: $\Delta = 0.1$, $0.5$ and $0.75$.
As shown in Fig.~\ref{fig:Ratioeq_0.1}(b), Fig.~\ref{fig:Ratioeq_0.5}(b), and Fig.~\ref{fig:Ratioeq_0.75}(b),
the QSH correlation ratio $R^{QSH}$ increases toward $1$ at intermediate values of $\lambda$,
indicating a robust QSH state for all three cases.   We have  checked  explicitly  that the   eigenvector corresponding to  the largest 
eigenvalue  matches  the  sign structure $\nu_{\ve{i}\ve{j}} $ of the Kane-Mele model.     Furthermore,  in Appendix~\ref{App:Flux} we  have  used  the 
flux-insertion scheme  presented in Ref.~\onlinecite{Assaad12}   to  probe for  the  topological invariant. 
The DSM-QSH and QSH-SSC phase boundaries are estimated by fitting the equal-time correlation ratios of the two largest lattice sizes $L=15$ and $L=18$  according to  the following function
\begin{equation}
f(L,\lambda)=R_c+a_1(\lambda-\lambda_c)L^{1/\nu}+a_2(\lambda-\lambda_c)^2L^{2/\nu}.
\end{equation}
The details of the fit are listed in Tab.~\ref{tab:DSM-QSH} and Tab.~\ref{tab:QSH-SSC}.
Remarkably, the critical values of $\lambda$ where superconducting order develops, as shown by the crossing points of $R^{SSC}$ in
Fig.~\ref{fig:Ratioeq_0.1}(a), Fig.~\ref{fig:Ratioeq_0.5}(a), and Fig.~\ref{fig:Ratioeq_0.75}(a), match the values of $\lambda$ where the QSH order vanishes. This indicates that, regardless of the strength of anisotropy,  direct phase transitions exist between the QSH and SSC phases.
The order parameters give consistent results, as shown in Fig.~\ref{fig:m}.

The comparison of the mean-field, Fig.~\ref{fig:MF_Diagram},  and   QMC, Fig.~\ref{fig:Phase_diagram},    phase  diagrams  are  very  instructive.  
The  transition from the  DSM to  QSH  insulator  at  $\Delta<1$ belongs  to the U(1)  Gross Neveu  universality  class.   The  essence of this transition,  a  symmetry-breaking  induced   electronic  mass generation,  is  captured  at the mean-field  level.  In fact,  an  $\epsilon$-expansion  around  the upper  critical  dimension accounts rather well  for  this  transition   for the  SO(3)   \cite{Assaad13, Liu21} and   U(1)  cases \cite{Otsuka18}.    
It is hence not   unexpected  that the  comparison  between the mean-field,  Fig.~\ref{fig:MF_Diagram},  and  QMC, Fig.~\ref{fig:Phase_diagram},    phase  diagrams  is  \textit{good}  
for this  transition.  In  contrast,  the  competition and interplay  between  the  QSH  and SSC phases are  radically  different  at  the mean-field  and  QMC levels.  We   interpret  this mismatch  as  a hint   that topology -- not  accounted for  at the mean-field level -- is  crucial  for  the 
understanding of the intertwinement of  the  QSH and SSC phases.  Of  particular  importance  is  that the QMC  phase diagram does not  show a  coexistence of the QSH and  SSC  phases.  The nature of the transition will be discussed in the next section.

\begin{table}[htp]
\caption{\label{tab:DSM-QSH}
    DSM-QSH crossing points $\lambda_c$.}
	\centering
	\begin{tabular}{cccc}
		\toprule
		\hline
		anisotropy&$\lambda_c$ &$\chi_r^2$&  $O$ \\
		\midrule
		\hline
		$\Delta=0.1$ & 0.0243(2)   & 0.14  & QSH \\
		$\Delta=0.5$ & 0.0225(2)   & 2.7  & QSH\\
		$\Delta=0.75$ & 0.0208(3)  & 0.81  & QSH\\
		 \hline
        \bottomrule
	\end{tabular}
\end{table}

\begin{table}[htp]
\caption{\label{tab:QSH-SSC}
    QSH-SSC crossing points $\lambda_c$.}
	\centering
	\begin{tabular}{cccc}
		\toprule
		\hline
		anisotropy&$\lambda_c$ & $\chi_r^2$&   $O$ \\
		\midrule
		\hline
		$\Delta=0.1$ & 0.06006(8)  &  0.08 & QSH \\
		 & 0.05988(2)  & 2.41 &   SSC \\
		$\Delta=0.5$ &0.0448(2) & 19.87 &   QSH\\
		& 0.04444(4)  &3.77 &   SSC \\
		$\Delta=0.75$ & 0.03829(4) &  1.67  & QSH\\
		 & 0.03788(3)  &  3.44&   SSC \\
        \hline
        \bottomrule
	\end{tabular}
\end{table}

Our QSH insulator at zero temperature is also a gapless phase,
reflecting the emergence of Goldstone modes upon  breaking  the  global  XY symmetry.  Therefore, both the spin current
operators $ \hat{J}^{x,y}$ and the corresponding angular momentum operator $ \hat{S}^z $ reveal gapless excitations around zero momentum.  
On the other hand,  merons in this phase   
are another low-energy excitation with 
a finite gap.  Roughly speaking, 
the binding energy of pairs of merons,  
is higher in the case of strong 
anisotropy. 
The  numerical  evidence  that merons bind    can be  deduced  from the   spectral  functions 
  presented  in  Appendix \ref{App:Spectrum}.   A comparison  between  the  single particle    and   superconducting 
spectral  functions  shows  that  the  cost  of  adding   a  pair  is  less  than  twice  the  single-particle  gap. 
This binding energy 
can be tuned to zero by increasing the interaction,  thus    triggering  a 
direct   transition  to a  superconducting  state.

\subsection{Nature of the QSH-SSC phase transition}
\label{sec:Transition}

Our most important result is the  seemingly continuous   nature  of  the QSH-SSC transition.

\begin{figure}[htbp]
\centering
\includegraphics[width=0.45\textwidth]{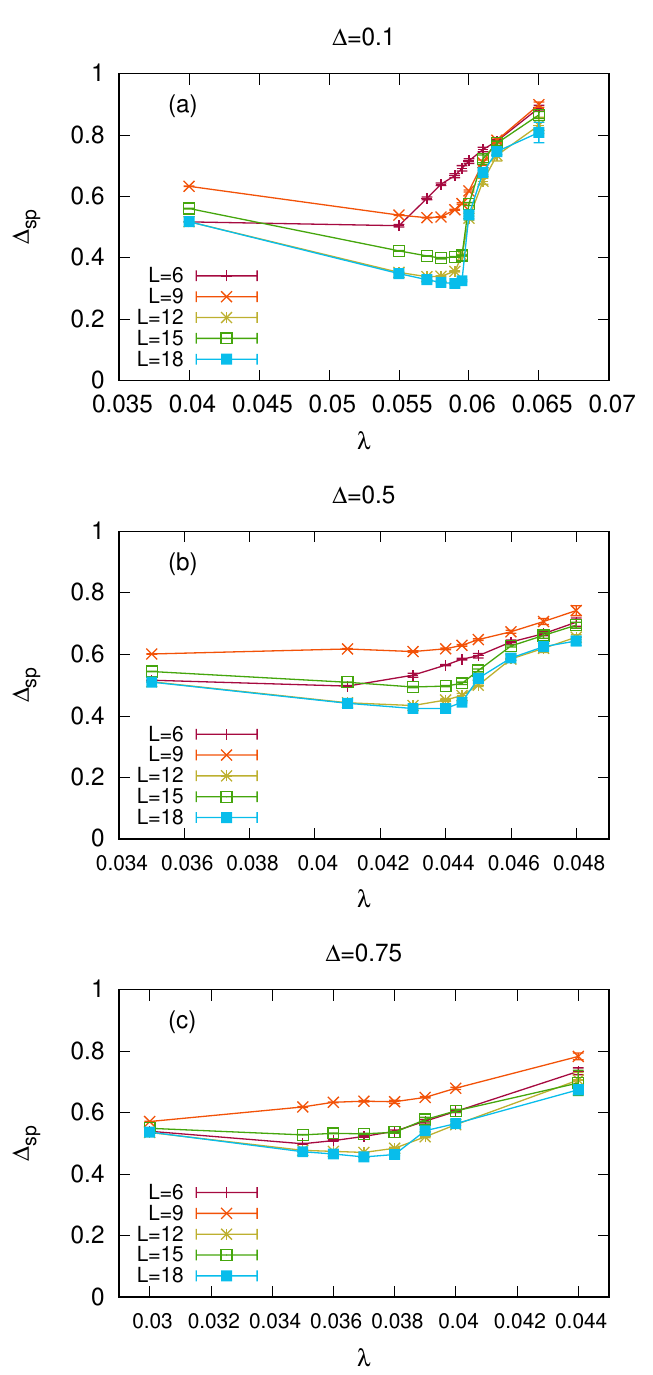}
\caption{\label{fig:sp_lamb}
Single particle gap $\Delta_{sp}$ across the QSH-SSC transition for $\Delta=0.1$ (a), $\Delta=0.5$ (b), and $\Delta=0.75$ (c), respectively.
}
\end{figure}

The QSH-SSC transition  has only bosonic excitations at low energy.  
To characterize  this,  we extrapolate the fermionic single particle gap $\Delta_{sp}$ from Green's  function:  
\begin{equation}
\begin{aligned}
   \sum_{\sigma} \langle  c_{\bm{k},\sigma } (\tau)  c^{\dagger}_{ \bm{k}, \sigma }(0) \rangle  \propto  e^{ -\Delta_{ \text{sp} } \tau }, 
   \end{aligned}
   \end{equation}
   at $ \bm{k} = M \equiv (0,\frac{2 \pi}{ \sqrt{3}}).$
Figure.~\ref{fig:sp_lamb} demonstrates that $\Delta_{sp}$ remains nonzero across the QSH-SSC transition at $\lambda_{c}$ for all three
considered  values of   the anisotropy.

In the following step, we inquire whether the QSH-SSC transition is  continuous  or  not. 
Considering that the computational cost of the AFQMC scales as $ \theta L^3 $ with $\theta$ being the projection length, we do not apply the commonly used method  for detecting first-order transitions, such as analyzing the finite-size behavior of the histogram of order parameters or the behavior of the Binder cumulant.
 Instead, we study the correlation length to reflect the nature of the  phase transitions.
 A continuous phase transition is characterized by a diverging correlation length in the thermodynamic limit:  $\xi \propto |\lambda-\lambda_c|^{-\nu}$. On the other hand, for a first-order transition, the correlation length saturates to a finite value.
 We use  the real-space, equal-time correlation functions $S^{O}(\bm{r})$ of the order parameter  to define the correlation length\cite{Toldin14}
 \begin{equation}\label{Eq:def_xi}
 {\xi}^{O} \equiv \sqrt{ \frac{ \sum_{\boldsymbol{r}}  | \boldsymbol{r} |^2  S^{O}(\boldsymbol{r})  }{ \sum_{ \boldsymbol{r}}  S^{O}(\boldsymbol{r}) }}
\end{equation}
where $S^{O}(\bm{r})$ is defined in Eq.~\ref{Eq:def_S_r}. 
 For  continuous symmetry  breaking,   an  issue with this definition is that it  picks  up the   correlation  
lengths along both the longitudinal and  the transverse directions.    
Hence without a specific symmetry-breaking pinning field to resolve the
longitudinal direction, the correlation length $\xi$ in Eq.~(\ref{Eq:def_xi})
is well defined only in the disordered state or at the critical point.
Therefore, we discard the data for the QSH (SSC) correlation length in  the QSH (SSC) 
state.

\begin{figure}[htbp]
\centering
\includegraphics[width=0.45\textwidth]{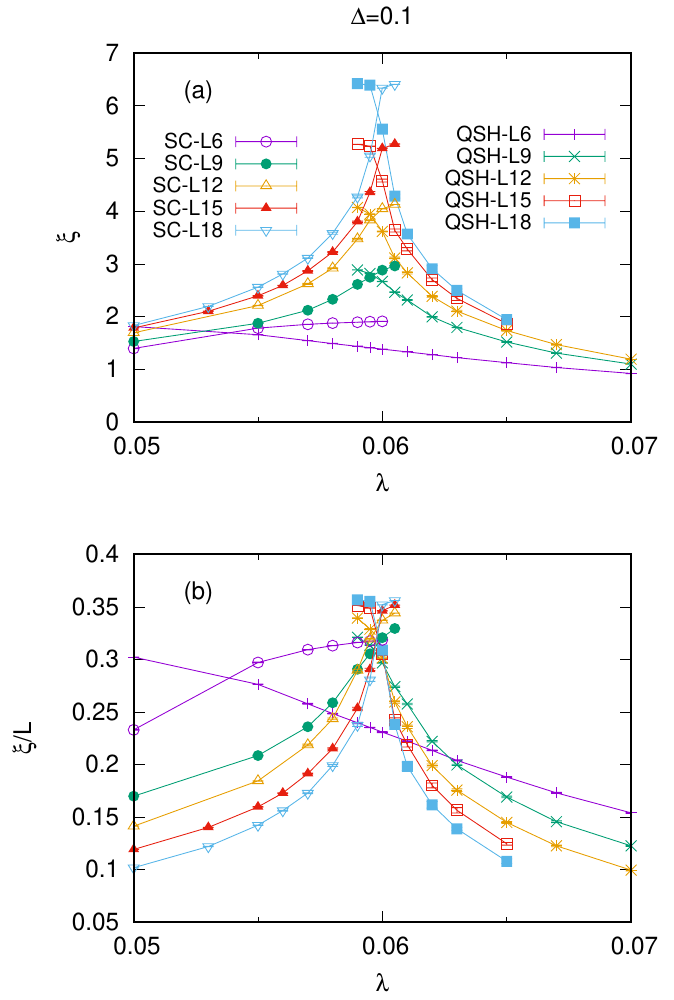}
\caption{\label{fig:cl0.1}
Correlation length (a) and scaled correlation length (b) in disordered phase for $\Delta=0.1$.
}
\end{figure}

\begin{figure}[htbp]
\centering
\includegraphics[width=0.45\textwidth]{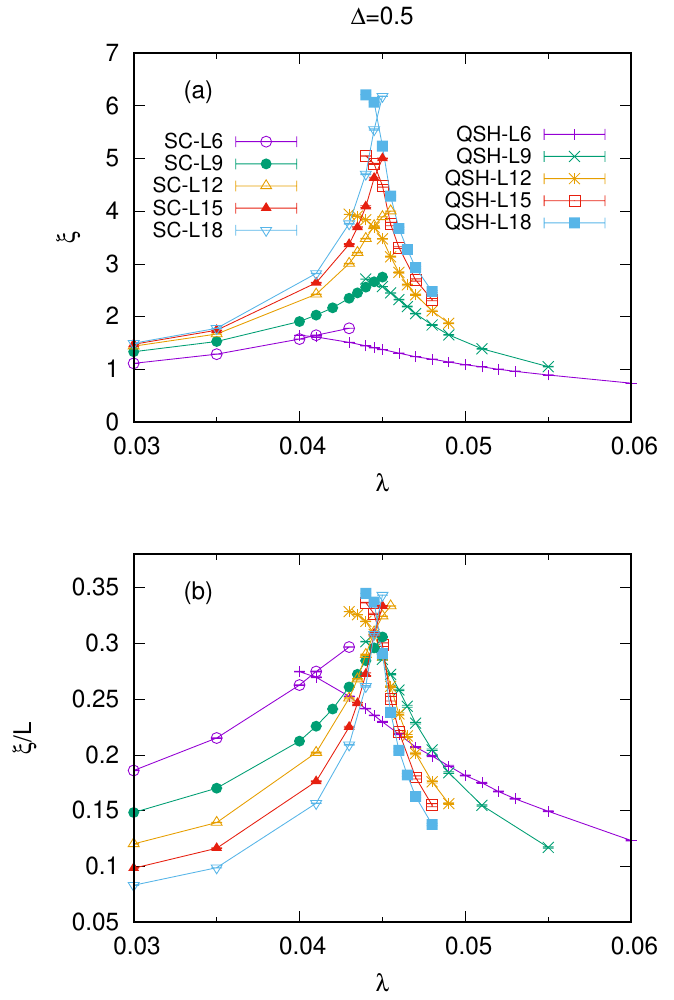}
\caption{\label{fig:cl0.5}
Same as Fig.~\ref{fig:cl0.1} for $\Delta=0.5$.
}
\end{figure}

\begin{figure}[htbp]
\centering
\includegraphics[width=0.45\textwidth]{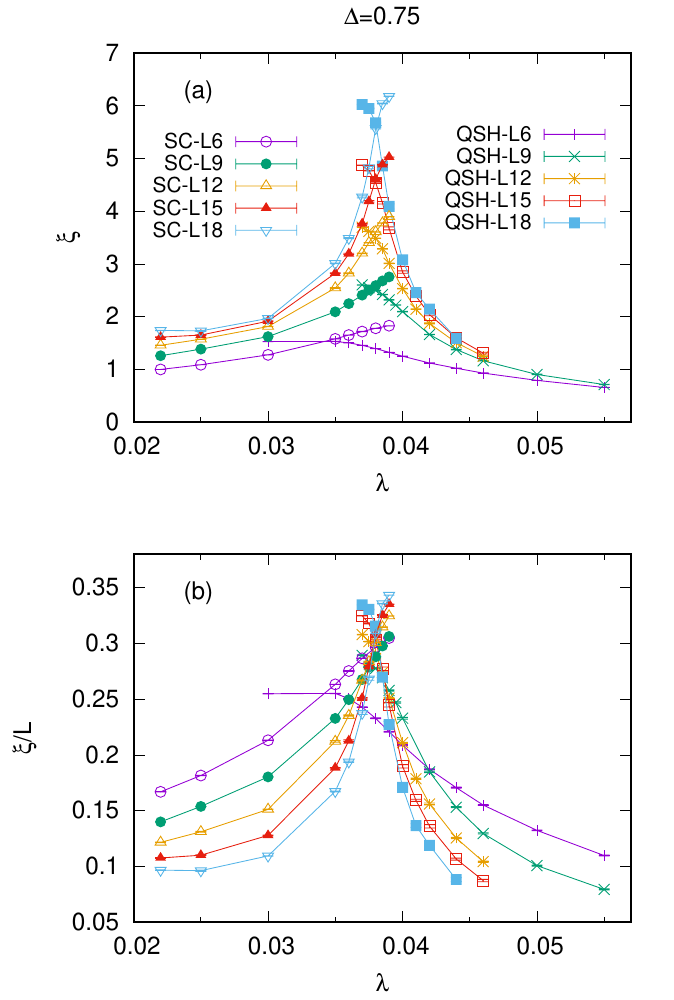}
\caption{\label{fig:cl0.75}
Same as Fig.~\ref{fig:cl0.1} for $\Delta=0.75$.
}
\end{figure}

 As depicted in Fig.~\ref{fig:cl0.1}(a), Fig.~\ref{fig:cl0.5}(a) and Fig.~\ref{fig:cl0.75}(a), around transition points at different values of $\Delta$,
 the correlation length $\xi$ of both the  QSH and SSC order parameters
 grow with system size $L$,    
 without any tendency of
 saturation.
 We define the scaled correlation length $\xi/L$ as the ratio between the correlation length
 and system size. As shown in Fig.~\ref{fig:cl0.1}(b), Fig.~\ref{fig:cl0.5}(b), and Fig.~\ref{fig:cl0.75}(b),
 for both the  QSH and SSC correlation lengths, the ratios for different system sizes cross
 at the same point, suggesting that the correlation length diverges with $L$. 
 The divergence of  correlation lengths indicates  the 
 continuous nature of the phase transition.

 It is worth mentioning that
 we observe an amazing match of the value of the scaled correlation length $\xi/L$   for 
  different anisotropy strengths $\Delta$.   
  The fact that the same value of $\xi/L$ at the three transition points implies that all 
 three QSH-SSC phase transition points correspond to the same  fixed point. 
 $\xi/L$ is a dimensionless quantity which is a renormalization-group invariant at the critical point. 
This number is claimed to be universal in a $(2+1)$ dimensional system  
 with conformal invariance.  On a lattice system, this universal number is not only pinned by 
 the scaling dimension of the order parameter but also by the microscopic couplings in 
 different directions, the boundary conditions, and the shape of the system (e.g. the aspect ratio).~\cite{Kamieniarz_1993, PhysRevE.94.052103}  
In our case, the only difference between the three different values of $\Delta$     
is the intrinsic spin anisotropy which is not related to the lattice geometry,  
such that $\xi/L$ should be universal   
if all three transitions  at $\Delta=0.1,0.5$ and $0.75$ belong to the same universality class.  
We also observe the interesting behavior that QSH and SSC operators cross at the same value of   
$\xi/L$ at the  transition point.  This also indicates  the identical value of  the 
anomalous dimension $\eta$ between the two order parameters, which is significant evidence of  an emergent O(4) symmetry~\cite{WangC17, SatoT17}.


We also calculate the first-order partial derivative of the free energy  density with respect to the coupling strength $\lambda$,
\begin{equation}
   \frac{\partial f}{\partial\lambda} =  \frac{1}{L^2}\frac{1}{\lambda} \langle \hat{H}_{\lambda} \rangle,
\end{equation}
to study the nature of the  phase transition.  This approach requires no information of the order parameter (and the associated symmetry breaking).    
In the case of a first order transition, when the system size is much larger than     
correlation length $\xi$,   
one expects  a   discontinuity in this derivative at  the transition point.   
\begin{figure}[htbp]
\centering
\includegraphics[width=0.45\textwidth]{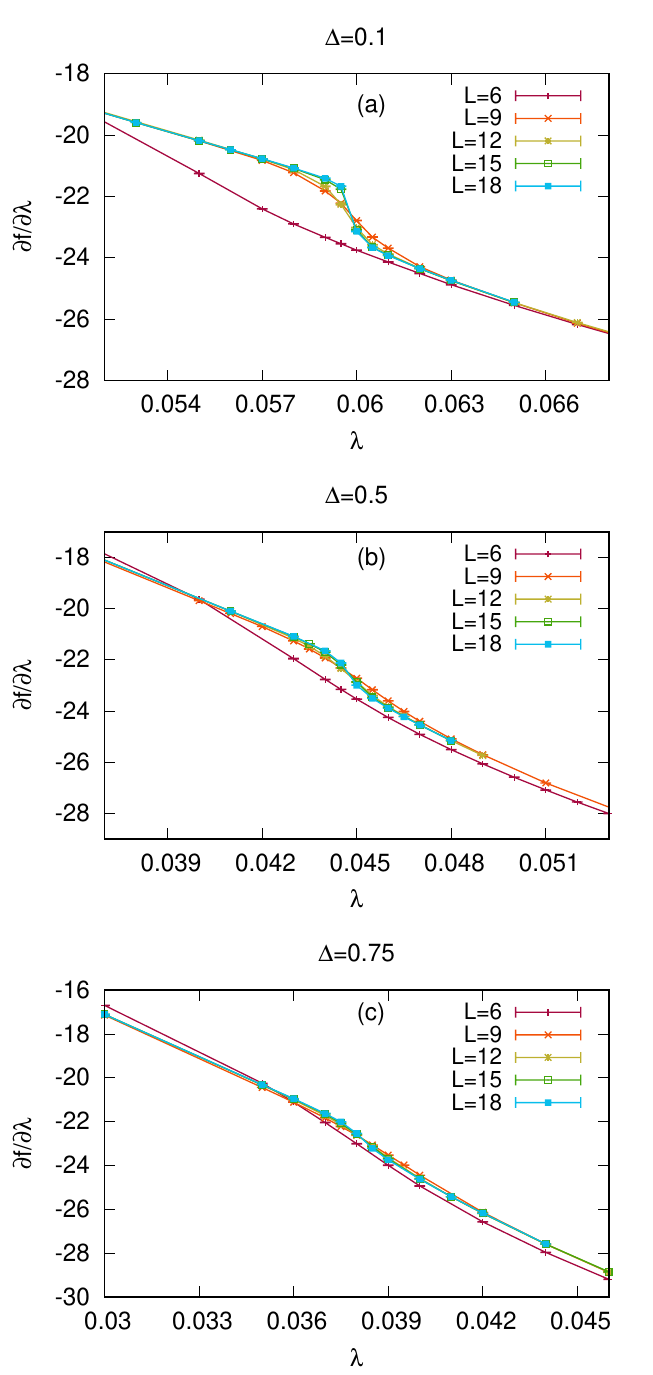}
\caption{\label{fig:PFPL}
Free-energy derivative ${\partial f/\partial\lambda}$ as a function of $\lambda$ in the vicinity of the QSH-SSC transition point,
for $\Delta=0.1$ (a), $\Delta=0.5$ (b), and $\Delta=0.75$ (c), respectively.
}
\end{figure}
Figure~\ref{fig:PFPL} shows $\partial f / \partial \lambda$ as a  function of $\lambda$ in the vicinity of the QSH-SSC transition 
point for $\Delta=0.1$, $0.5$ and $0.75$.
Our data reveal no clear signs of a  jump for  the  accessible  system sizes in this study. This is
consistent with the aforementioned correlation length analyses.

In the strong anisotropic case $\Delta=0.1$,
the slope of the curve scales up 
moving towards the transition point when increasing
the system size.  
This  result may be interpreted as the signature of a `weakly first-order' transition.
Around a continuous transition point, the derivative of the free energy scales as
\begin{equation}
    \frac{\partial f}{ \partial \lambda} \propto
    | \lambda - \lambda_c |^{ (d + z) \nu - 1 }
\end{equation}
in the thermodynamic limit.  For a `weakly first-order' transition,
one  expects a `pseudo critical' phenomenon where
the $\nu(L)$ estimated from finite sizes would approach $1/(d+z)$ as $L$ reaches $\xi$,\cite{SandvikJPSJ2019} 
such that $ \partial f / \partial \lambda$
asymptotically shows a jump. 
However, we won't be able to conclude the nature of transition at $\Delta=0.1$  
since it is not clear  whether the finite-size slope diverges or saturates upon approaching 
the thermodynamic limit.  
On the other hand, the robustness of the slope in the case of $\Delta=0.5$ and $\Delta=0.75$ as shown in Fig.~\ref{fig:PFPL}(b) and (c), 
indicates clear continuous phase transitions,  unless there exists a non-diverging $\xi$ that is significantly larger than $L$.

\begin{figure}[htbp]
\centering
\includegraphics[width=0.45\textwidth]{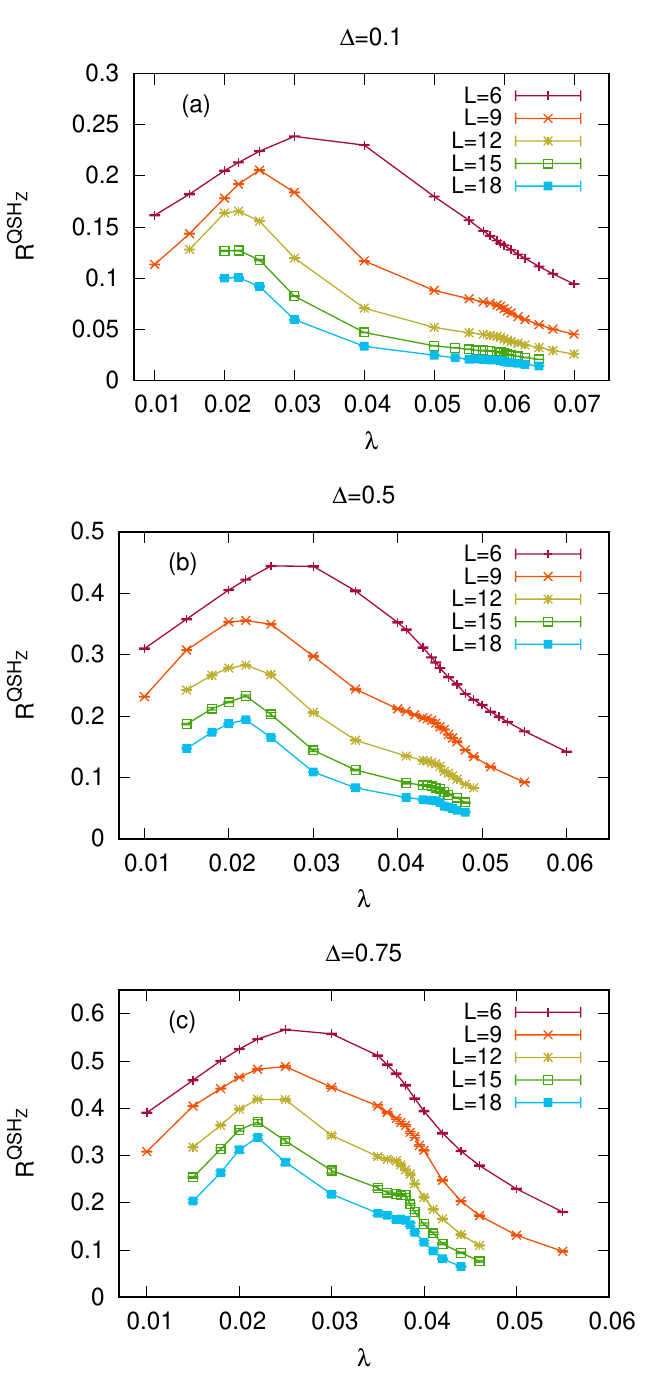}
\caption{\label{fig:QSH_Zeq}
Equal-time correlation ratio $R^{QSHz}$ for $\Delta=0.1$ (a), $\Delta=0.5$ (b), and $\Delta=0.75$ (c), respectively.
}
\end{figure}

We  now  consider  the  Z-component of  the  QSH  correlations, Eq.~\ref{Eq:def_qshZ}.  
In the DSM  and  at  the   DSM  to  QSH  transition   the  single  particle  gap  vanishes  such  that  this  quantity  is  expected 
to  decay as a  power-law.  In particular,  in the  DSM  the  scaling dimension of  the fermion operator  is 
given  by  $\frac{d} {2} $   ($d=2$ is  the dimensionality)  such  that the  Z-QSH  as  well as   XY-QSH  correlation  functions  are  
expected to  decay  as    $r^{-4}$  with $r$  being the  distance.  At  the  DSM to  QSH   transition  the scaling   dimensions  of  both   
 quantities will differ.    For  equal  time   correlations,   and in  $d=2$,    power-law  decay  leads  to a   divergence  in  the  structure  factor   provided  that  the  scaling  dimension of the 
operator  is  smaller  that   unity.   In the  DSM  the  scaling    dimension of the QSH operators is  two,  and  we do  not  pick  up any a  signal: 
as  apparent from    Fig.~\ref{fig:QSH_Zeq},  the correlation ratio  decays  as  a  function of  system size. 
 However  at  the  DSM-QSH transition  it  
is  worth noting    a  distinct  cusp  in  the  $S^{\text{QSH}_Z} $    correlation  ratio  especially   at   $\Delta  =  0.75$  (see   Fig.~\ref{fig:QSH_Zeq}(c)). 
In the  ordered  XY-QSH  phase,  the   anisotropy  opens  a  gap   in the  Z-QSH   spectrum,  leading to a reduction in the Z-QSH  correlation  
ratio as  a  function of  system  size  (see   Fig.~\ref{fig:QSH_Zeq}).     In the  SSC  phase the  spin   degrees  of   freedom  are 
gapped,   such   that    the 
Z-QSH  correlations  decay   exponentially.   Again, in this  phase,   the  Z-QSH  correlation   ratio   decays    as  a  function of  system size  
 (see   Fig.~\ref{fig:QSH_Zeq}).

 It  hence  comes  as a    surprise  that  at  the QSH-SSC  transition,   we  see  a  distinct  cusp in  the  Z-QSH correlation  ratio. This suggests  that at this  transition 
 \begin{equation}
  S^{\text{QSH}}   (\bm{Q})  \propto  L^{ 1 - \eta_{ X Y } }
\end{equation}  
and
\begin{equation}
 S^{\text{QSH}_Z} (\bm{Q})  \propto  L^{ 1 - \eta_{ Z } }
\end{equation}
albeit  with different  scaling  dimensions. 
Here,  $\bm{Q}\equiv(0, 0)$.
 Assuming  the   same  criticality  as  in Ref.~\onlinecite{Qin17},  $\eta_{XY} \approx 0.13 $ and $\eta_{Z} \approx 0.91$. 
Given  the  large  anomalous  dimension,  $\eta_Z$, it  is  more  advantageous  to    consider   the \textit{susceptibility} as defined in Eq.~\ref{Eq:def_sus_Z}  of   Appendix \ref{App:Time}.    For  Lorentz  invariant  systems,  this  quantity  scales  as   $L^{2 -  \eta_{Z}} $  and  
suffers   less from  background  effects.  Fig.~\ref{fig:QSH_Z}   plots  the  correlation  ratio  as  obtained  from  the   \textit{susceptibility}.   
While   we  can  observe  clear  cusps  at  the   QSH-SSC transition   we  cannot   unambiguously  claim  that  this  quantity  scales  to 
a  finite value in the  thermodynamic limit.

\begin{figure}[htbp] 
\centering
\includegraphics[width=0.45\textwidth]{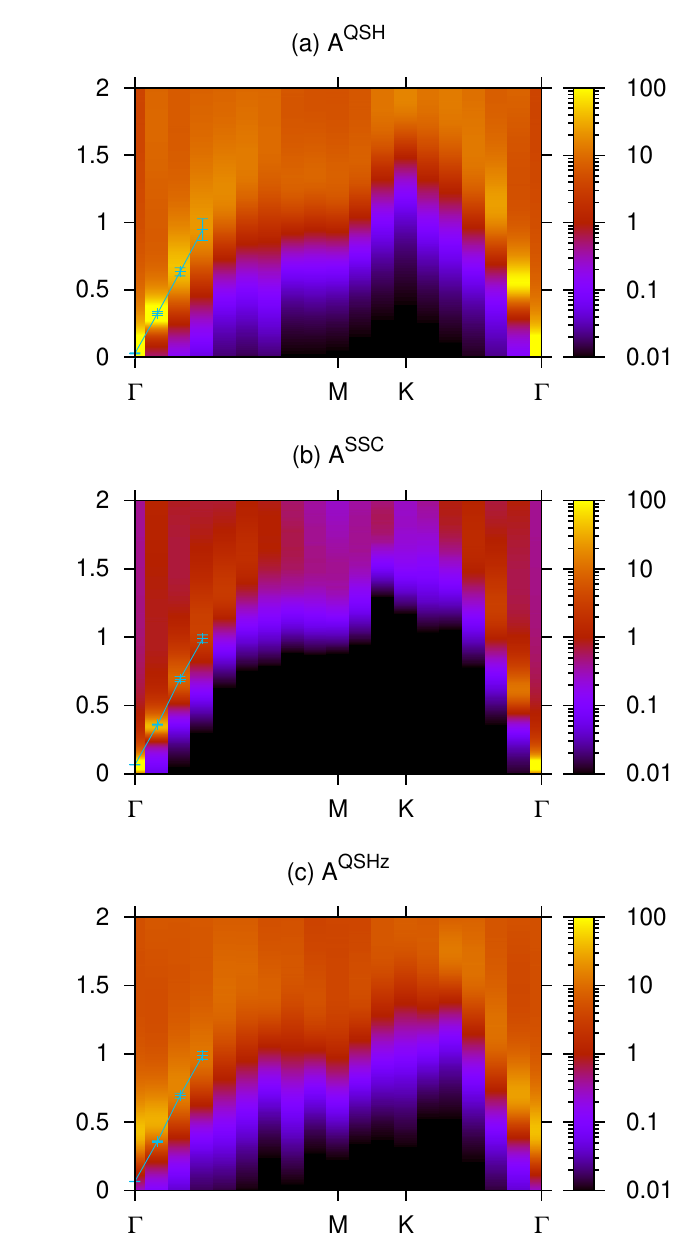} 
\caption{\label{fig:SW_QCP} 
QSH (a), SSC (b), and QSHz (c) dynamical spectrum at the  QSH-SSC critical point ($\lambda=0.038$)  
for $\Delta=0.75$. 
Here, $L=18$. Blue lines in (a) and (b) 
are the momentum dependence of the extrapolated excitation gap of 
$ \hat{J}^{XY} $ and $ \hat{\eta}^+(\hat{\eta}^{-} ) $  
operators. The blue line in (c)  is copied from the SSC dispersion relation in (b) to guide the eye. 
}
\end{figure}

In  Fig.~\ref{fig:SW_QCP}  we show    
the  spin-orbit, $ \hat{J}^{XY} $, and  the  pairing, $\hat{\eta}^+$($\hat{\eta}^-$),  dynamical  
correlation  functions at  the  QSH-SSC   critical point  at  $\Delta  = 0.75$.   For  the  definition of the spectral 
functions   as  well as  for  further  data, we  refer  the  reader  to  Appendix \ref{App:Spectrum}. 
The  spectral  functions  display gapless excitations with  the   very  same  velocity. 
This  stands in accord  with   emergent  Lorentz invariance.      
The  $\hat{J}^Z$  spectral  function, Fig.~\ref{fig:SW_QCP} (c),  allows  for   an  interpretation in terms of 
   gapless  excitations    at   the 
$\Gamma$ point   albeit  with  small  spectral  weight  at  low  energies.

\section{Discussions and outlook} 
\label{Sec:Conclusions}

Our  model  realizes  an    easy-plane  quantum spin Hall  insulating  state that emerges from spontaneous  U(1)  broken  spin  symmetry,   with  a  unnormalized three 
component  order parameter $\ve{N}$  defined  in   space-time.     The  norm of  the order  parameter  defines the  single-particle 
gap.       By  tuning a  single  parameter $\lambda$  at  a  given anisotropy  strength,    we   observe    DSM-QSH  as  well as  QSH-SSC   transitions. 
 At the  DSM-QSH  transition,   the  amplitude  of  the order parameter  vanishes  and   the  single-particle  gap  closes.   We  understand  this transition  in terms of    Gross-Neveu-XY  universality,   the  exponents  of  which  should  be  equivalent to  those   computed  in   Ref.~\onlinecite{Otsuka18}.  

The  focus of  the paper,   is  on the  QSH-SSC  transition.  Here, the  QMC results   show that  i) the  single  particle  gap  does  not  close  at    the  
transition and  ii)   in  contrast  to  mean-field  calculations,   our  results on lattice  sizes up  to  $L=18$  support  a  continuous and 
 direct  transition   for all considered  values of  the  anisotropy.       We  can  safely   omit   the  scenario  of  fine  tuning, where  accidentally, the  two 
 transitions occur  at the same  value of  $\lambda$.    This statement is based  on the numerical observation  that  we consistently see   a  direct  
 transition for  all values of  the  anisotropy.  Furthermore,    the energetics  are affected  by  the value  of   the imaginary time  step we  use in our simulations.  Were  we  at a  fine  tuning point,  then  we  would have noticed  substantial  changes in our results  when  varying the  
 imaginary time step. 
 
 Since  the  single  particle  gap  remains  finite  we    can normalize  the 
 order parameter  vector  $\ve{n} =  \ve{N}/|\ve{N}|$, and   attempt to  understand the  transition in terms of  fluctuations of  $\ve{n}$.  In this  
 context,  and  as discussed  in the  introduction,  skyrmions or   pairs of merons  carry   charge  2e  and  condense  at  the transition.    
This suggests  that our QSH-SSC transitions flow to the easy-plane deconfined 
quantum critical point irrespective of the  anisotropy  parameter. 
This is also supported numerically by  the  observed  \textit {universal} value of  
the  scaled correlation length upon changing  the 
anisotropy.  
The scaled correlation length  takes the same value for  the QSH and SSC   fluctuations at  the   easy-plane DQCP   
which  supports an emergent O(4) symmetry  at  least on  intermediate  length  scales. 
Our model  summarizes  the  very  first  calculations of  this critical point in a lattice Hamiltonian  
where  the lattice   regularization does not  break the  IR  symmetries  of  the  
putative  field  theory.  As a consequence, our lattice  regularization does not  introduce 
quadruple monopoles as in  the  easy-plane   JQ model.     We  refer  the  reader  to  Appendix   \ref{monoploes.sec}      for a 
detailed  discussion of  this point.   

One of  the  characteristics of  the  easy-plane DQCP,     is  the  emergence  of  deconfined  spinons  at the  critical  point.   
In particular,  one  can   adopt  a    CP$^1$  representation of  the order  parameter  
\begin{equation}
    \ve{n}  \equiv  z^{\dagger}  \ve{\sigma}  z   \quad \quad \quad z^\dagger z = 1
\end{equation}   
(see Appendix   \ref{monoploes.sec}).  The  claim   of   DQC \cite{Senthil04_2}  is  that   the 
CP$^1$   theory  supports  a   deconfined phase at  the critical point.    The  questions then becomes 
how  we  can  provide  numerical evidence for  this. 
Fractionalized spinons are not directly measurable at  the DQCP point since they do not
directly correspond to any local second quantized operators.   However,   the  existence  of   deconfined 
spinons  suggests  that at   criticality  the  Z-component   of  the  correlation functions of the  QSH 
order parameter  shows  power-law  decay  since   it  just corresponds   to   spinon correlation  functions. 
Our  numerics  support this point of  view.

To place  our  results in a   broader  perspective,  we  can  ask  the question of whether 
designer Hamiltonians
with higher symmetry  can impact  criticality. 
Although it is well known that phase transitions numerically observed in easy-plane 
lattice spin models   have a higher tendency to be discontinuous in the  anisotropic case,   
two possible underlying physical interpretations  remain possible. 
First, the relevance of the $Z_4$ symmetry-breaking     
perturbation at the easy-plane deconfined fixed point generally leads to a runaway flow, 
explaining   the first order nature of transition.    
Second, even if symmetry-breaking terms due to lattice regularization (e.g., $O(4)$ down to $U(1) \times U(1)$ 
or $U(1) \times Z_4$) are  imposed to be zero,  the easy-plane DQCP may not even exist in any unitary conformal field theory.  
In this  context   our  results may  be  understood  in  terms  of  proximity  to a   DQCP  that  is not  accessible  to   our 
simulation  space.  This  could  correspond  to  a  DQCP  in the complex  plane \cite{WangC17}  or  in dimensions  close  to d=2 \cite{Nahum19,WangC19}.

\ \\ 
\subsection*{
ACKNOWLEDGMENTS}
 The authors gratefully acknowledge
the Gauss Centre for Supercomputing e.V. for funding this
project by providing computing time on the GCS Supercomputer SUPERMUC-NG at Leibniz Supercomputing Centre.
F.F.A. acknowledges the  DFG  for   funding  via  W\"urzburg-Dresden Cluster of 
Excellence on Complexity and Topology in Quantum Matter ct.qmat (EXC 2147, Project ID 390858490)  as  well  as 
the SFB1170 on Topological and Correlated Electronics at Surfaces and Interfaces.
 T.S. acknowledges funding from the Deutsche Forschungsgemeinschaft under Grant No. SA 3986/1-1. Y.L.
was supported by the China Postdoctoral Science Foundation
under Grants No. 2019M660432 and No. 2020T130046 as
well as the National Natural Science Foundation of China
under Grants No. 11947232 and No.U1930402. D.H. and W.G. were
supported by the National Natural Science Foundation of
China under Grants No. 12175015 and No. 11734002.

\bibliography{zwang}

\clearpage

\appendix
\section{Mean field calculation}
\label{Appendix}

In this appendix, we present  our  mean-field calculation. Expanding Eq.~\ref{Eq:Ham_V} of the main text as

\begin{equation}
\begin{aligned}\label{Eq:MF_decouple}
 H_\lambda = & -\lambda \sum_{\varhexagon}   
   \left[  \left( \sum_{\langle \langle \bm{i} \bm{j} \rangle \rangle  \in \varhexagon }   \hat{J}^x_{\bm{i},\bm{j}} \right)^2  
   +    \left( \sum_{\langle \langle \bm{i} \bm{j} \rangle \rangle  \in \varhexagon }   \hat{J}^y_{\bm{i},\bm{j}} \right)^2     \right.      \\  
 &  \phantom{=\;\;} \left. 
 +  \Delta  \left( \sum_{\langle \langle \bm{i} \bm{j} \rangle \rangle  \in \varhexagon }   \hat{J}^z_{\bm{i},\bm{j}} \right)^2     \right]\\
   = &  -\lambda  \sum_{\varhexagon}  \sum_{ \langle \langle \bm{i} \bm{j} \rangle \rangle }
   \sum_{ \langle \langle \bm{i'} \bm{j'} \rangle \rangle \neq  \langle  \langle \bm{i} \bm{j} \rangle \rangle  }
  \bm{ \hat{J}^x_{ \langle \langle i, j \rangle \rangle }  \cdot  \hat{J}^x_{ \langle \langle i', j' \rangle \rangle }  } \\
  & + \bm{ \hat{J}^y_{ \langle \langle i, j \rangle \rangle }  \cdot  \hat{J}^y_{ \langle \langle i', j' \rangle \rangle }  } + \Delta \bm{ \hat{J}^z_{ \langle \langle i, j \rangle \rangle }  \cdot  \hat{J}^z_{ \langle \langle i', j' \rangle \rangle }  } \\ 
   & - \lambda \sum_{\varhexagon}  \sum_{ \langle \langle \bm{i} \bm{j} \rangle \rangle }
   [ + (4+2\Delta) \hat{\eta}^{\dagger}_{\bm{i}}  \hat{\eta}_{\bm{j}} + h.c  +...]
\end{aligned}
\end{equation}
where
\begin{equation}
\begin{aligned}
 &  \bm{ \hat{J} }^\alpha_{ \langle \langle i, j \rangle \rangle }  \equiv    i \nu_{ \bm{i} \bm{j} }
  \hat{c}^{\dagger}_{\bm{i}} \bm{\sigma}^\alpha \hat{c}_{\bm{j}}  + H.c.,         \\
 &  \hat{\eta}_{\bm{i}}  \equiv  \hat{c}_{ \bm{i} \downarrow}  \hat{c}_{ \bm{i} \uparrow},     \   \   \    \
    \hat{\eta}^{\dagger}_{\bm{i}}  \equiv  \hat{c}^{\dagger}_{ \bm{i} \uparrow}  \hat{c}^{\dagger}_{ \bm{i} \downarrow}.
\end{aligned}
\end{equation}
The ellipsis  denotes   terms that do not contribute to  the  SSC or QSH ordering   within the  mean-field decomposition.

The mean-field calculation involves selecting a polarization direction for the two components of the QSH  and SSC order parameters. 
The calculation is done by numerically minimizing the free energy in the space of the two order parameters.

\begin{figure}[H]
\centering
\includegraphics[width=0.4\textwidth]{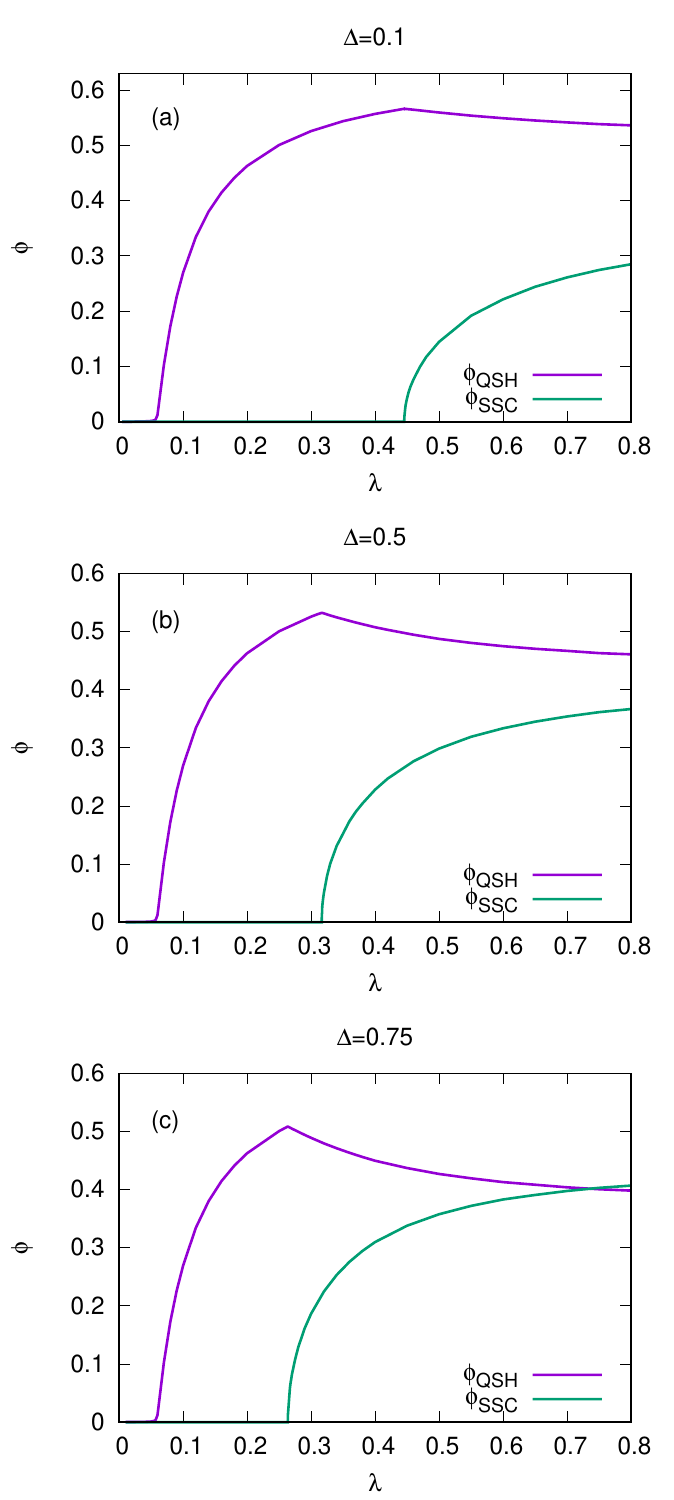}   
\caption{\label{fig:Mean_field_M}
Mean-field solution for the QSH and SSC order parameters as a function of $\lambda$ for $\Delta=0.1$ (a), $\Delta=0.5$ (b), and $\Delta=0.75$ (c), respectively.
}
\end{figure}

The two order parameters as a function of $\lambda$  for 
the half-filled case are shown in Fig.~\ref{fig:Mean_field_M}. For all three different values 
of anisotropy, we observe a Dirac semi-metal ($\phi_{QSH} = \phi_{SSC} = 0$), a pure QSH
state ($\phi_{QSH} \neq 0, \phi_{SSC} = 0$) as well as coexistence of 
QSH and  SSC phases ($\phi_{QSH} \neq 0, \phi_{SSC} \neq 0$).

The mean-field phase diagram in  Fig.~\ref{fig:MF_Diagram}    shows greater  stability of  the    QSH  phase  at  stronger 
 anisotropy.  The reason for  this  becomes transparent when taking a glimpse  at  Eq.~\ref{Eq:MF_decouple}.  Here  $\Delta$  
 modulates  the magnitude  of 
 the  pair-hopping ($ \hat{\eta}^{\dagger}_{\bm{i}}  \hat{\eta}_{\bm{j}} + h.c .$.)  but not of  the  in-plane   spin-orbit   interactions   ( 
 $  \hat{J}^x_{ \langle \langle i, j \rangle \rangle }  \cdot  \hat{J}^x_{ \langle \langle i', j' \rangle \rangle}  +   x \leftrightarrow y $ ).

\clearpage
\section{Time displaced  observables}
\label{App:Time}

 To  define  susceptibilities  in the  realm  of  the zero  temperature  projective  QMC   algorithm used in this paper,  
we  distinguish  between  observables  $ \hat{O}_{\ve{q}}   = \frac{1}{\sqrt{N}} \sum_{\ve{r} } e^{i \ve{q}\cdot\ve{r}} \hat{O}_{\ve{r}} $ 
   that commute   or  not  
with the   Hamiltonian.    Here,  $N= L^2$. The   key point    is  that   for   $\left[  \hat{O}_{\ve{q}}, \hat{H} \right] =0 $,  
\begin{equation}
  \lim_{L \rightarrow \infty } \lim_{\beta \rightarrow \infty } \chi^{FT}(\ve{q},\beta,L) 
  \neq \lim_{\beta \rightarrow \infty } \lim_{L \rightarrow \infty } \chi^{FT}(\ve{q},\beta,L)
\end{equation}
where  
\begin{equation}
	\chi^{FT}(\ve{q},\beta,L)  =   \int_{0}^{\beta}  d  \tau  \left(   \langle   \hat{O}_{\ve{q}}(\tau)   \hat{O}_{-\ve{q}} \rangle_T  -   \langle   \hat{O}_{\ve{q}}   \rangle_T \langle \hat{O}_{-\ve{q}} \rangle_T \right).
\end{equation}
In the  above,   $ \langle  \bullet  \rangle_T  =   \frac{1}{Z} \text{Tr} \left( e^{- \beta  \hat{H}} \bullet  \right)   $  and   $  \hat{O}_{\ve{q}}(\tau)  =  
e^{\tau \hat{H}} \hat{O}  e^{-\tau  \hat{H}} $. In  particular,   if  the  ground  state is  not   degenerate   then  for   $\left[  \hat{O}_{\ve{q}}, \hat{H} \right] =0 $, 
$\lim_{L \rightarrow \infty }    \lim_{\beta \rightarrow \infty }  \chi^{FT}(\ve{q},\beta,L)   = 0 $,   while 
$    \lim_{\beta \rightarrow \infty }\lim_{L \rightarrow \infty }  \chi^{FT}(\ve{q},\beta,L)   $    does  not  necessarily vanish.   An 
 example is  the  spin susceptibility  for  a    tight-binding  model    when  the  boundary  conditions  are  
 chosen  to  ensure  that  the  ground state  is non-degenerate. 

Let us now consider  the  case  of  finite momentum $\ve{q} \neq 0 $   for  a momentum-conserving  Hamiltonian.   Hence,    $\left[  \hat{O}_{\ve{q}}, \hat{H} \right] \neq 0 $.   Provided  that the  ground state  is unique,  we   will show  that  the  limits  can  be interchanged. Our   starting point  is the Lehmann  representation: 
\begin{eqnarray}
    \chi^{FT}(\ve{q},\beta,L)    &&  =  \frac{\beta}{Z}  \sum_{n}    e^{-\beta  E_n} \left| \langle  n | \hat{O}_{\ve{q}} | n \rangle \right|^2   +  
       \\    && \frac{1}{Z} \sum_{n \ne m} \frac{e^{-\beta E_m} - e^{-\beta E_n}}{E_m - E_n} \left| \langle  n | \hat{O}_{\ve{q}} | m \rangle \right|^2  
       \nonumber 
\end{eqnarray}
with  $\hat{H} | n \rangle  = E_n  | n \rangle $  and  $n  \in  \mathbb{N} $.   Since we  have  assumed  that  $\ve{q} \ne  0 $,  the  first  term of  the   right-hand side  of  the  above equation  vanishes.     Defining   the  density  of  state  as,  $N(E) =    \lim_{L \rightarrow \infty } \sum_{n} \delta(E_n  -   E) $   we  obtain: 
\begin{eqnarray}	
	\lim_{L  \rightarrow \infty}  \lim_{\beta \rightarrow \infty} \chi^{FT}(\ve{q},\beta,L)   & & =  \int  d E  N(E)   {\cal P} \frac{1}{E-E_0}  \times \\ 
	 & & \left(   \left| \langle  E | \hat{O}_{\ve{q}} | E_0 \rangle \right|^2   +  \left| \langle  E_0 | \hat{O}_{\ve{q}} | E \rangle \right|^2 \right)   \nonumber 
\end{eqnarray}  
and  
\begin{equation}
   \lim_{\beta \rightarrow \infty}  \lim_{L  \rightarrow \infty}  \chi^{FT}(\ve{q},\beta,L)  =   N(E_0)
      \lim_{L  \rightarrow \infty}  \lim_{\beta \rightarrow \infty} \chi^{FT}(\ve{q},\beta,L)  
\end{equation}
For  a  unique   ground state,   $N(E_0) = 1$.  Hence,  under  the  aforementioned  assumptions  we  can interchange the limits  and  it  makes  sense  to 
define   susceptibilities  within the  ground state    algorithm,  where  we  first  take the limit  of  zero temperature  and  then  consider  larger and  larger  lattices.   

For practical purposes,  we  compute: 
\begin{equation}
   \chi ( \bm{q} ) = 
   \int_{0}^{\beta} \text{d} \tau    
    \langle \hat{O}_{\bm{q}} (\tau) \hat{O}_{-\ve{q} } (0)  \rangle  \\ 
  - \langle \hat{O}_{\bm{q}} (\tau) \rangle \langle \hat{O}_{- \bm{q} } (0)  \rangle 
\end{equation}
where  
\begin{equation}\label{Eq:def_PQMC}
    \langle \hat{O}_{\bm{q}} (\tau) \hat{O}_{-\bm{q}} (0)  \rangle  \equiv  
    \frac{ 
    \langle \Psi_T | e^{ -   \theta  \hat{H} } e^{ -( \beta -\tau ) \hat{H} }  \hat{O}_{\ve{q}}   e^{ -\tau \hat{H} } \hat{O}_{-\bm{q}}  e^{ -  \theta  \hat{H} }  | \Psi_T \rangle
    }{ 
    \langle  \Psi_T | e^{ -( 2\theta + \beta ) \hat{H} }  | \Psi_T \rangle }
\end{equation} 
and $| \Psi_T \rangle $ is the trial wave function.   
We consider $\beta=L$ for all three values of $\Delta$.  
As shown in the main text, 
we  implicitly checked  that for   the considered   size L,  
$\theta$ is  chosen to be large enough to converge to the ground state.

Since   the zero  temperature  approach  to  susceptibilities  matches   the  result  obtained   with  the  
\textit{traditional}     calculations,   the  scaling  behaviors  are  identical.  In particular  in the  vicinity of  a   
   Lorentz   invariant   (z=1)   critical point  we  expect:  
\begin{equation}
  \chi(\ve{q})       \propto  \xi^{-2 \Delta + d + 1 }
\end{equation} 
where  the   relationship   between the scaling   and  anomalous  dimensions reads   $  2 \Delta = d + z - 2 + \eta$  and  $\xi$  is  the diverging  
length  scale in space and time.      Replacing  the  length  scale  with the linear  size of our  system  yields  the   desired  result: 
\begin{equation}
	\chi(\ve{q})       \propto   L^{2 - \eta}.
\end{equation}  
Hence, as for the finite  temperature case,  we  expect at the critical point   that  $\chi$   
suppresses  background contributions of  the  non-singular part of free energy by an additional power (of the dynamical exponent). 
~ \cite{Liu18,Francesco}   
Assuming Lorentz invariance at the easy-plane DQCP,  this additional power  is unity.  

We define the susceptibilities of the QSH, SSC, and QSHz in \ref{Eq:def_sus_XY}, \ref{Eq:def_sus_Z} and \ref{Eq:def_sus_SSC}.
\begin{equation}
\begin{aligned}\label{Eq:def_sus_XY}
   \chi^{\text{QSH} }_{m, n} (\boldsymbol{q})
        \equiv &  \frac{1}{L^2} \sum_{\boldsymbol{r},\boldsymbol{r'}}   \int_{0}^{\beta} \text{d} \tau   e^{\mathrm{i}\boldsymbol{q}\cdot(\boldsymbol{r}-\boldsymbol{r}')} \\ 
    &    \langle  \hat{ O }^X_{\boldsymbol{r}, m}(\tau)  \hat{  O  }^X_{\boldsymbol{r'}, n }(0) 
    +  
     \hat{ O }^Y_{\boldsymbol{r},  m}(\tau)  \hat{ O }^Y_{\boldsymbol{r'},  n}(0) 
    \rangle.
\end{aligned} 
\end{equation}

\begin{equation}
\begin{aligned}\label{Eq:def_sus_Z} 
   \chi^{\text{QSH}_Z }_{m, n} (\boldsymbol{q})
        \equiv   \frac{1}{L^2} \sum_{\boldsymbol{r},\boldsymbol{r'}}   \int_{0}^{\beta} \text{d} \tau   e^{\mathrm{i}\boldsymbol{q}\cdot(\boldsymbol{r}-\boldsymbol{r}')}    
        \langle  \hat{ O }^Z_{\boldsymbol{r}, m}(\tau)  \hat{  O  }^Z_{\boldsymbol{r'}, n }(0)  
    \rangle.
\end{aligned} 
\end{equation}

\begin{equation}
\begin{aligned}\label{Eq:def_sus_SSC}
   \chi^{\text{SSC} }_{a, b} (\boldsymbol{q})
        \equiv  &  \frac{1}{L^2} \sum_{\boldsymbol{r},\boldsymbol{r'}}   \int_{0}^{\beta} \text{d} \tau   e^{\mathrm{i}\boldsymbol{q}\cdot(\boldsymbol{r}-\boldsymbol{r}')}     \\ 
	& [ \langle  \hat{ \eta }^{+}_{\boldsymbol{r},\boldsymbol{ \widetilde{\delta} }_a }(\tau)  \hat{ \eta  }^{-}_{\boldsymbol{r'},\boldsymbol{ \widetilde{ \delta} }_b }(0)  \rangle +  
	 \langle  \hat{ \eta }^{- }_{\boldsymbol{r},\boldsymbol{ \widetilde{\delta} }_a }(\tau)  \hat{ \eta  }^{+}_{\boldsymbol{r'},\boldsymbol{ \widetilde{\delta} }_b }(0)   
	\rangle],
\end{aligned} 
\end{equation} 

After diagonalizing the corresponding three  susceptibilities, we calculated the renormalization-group invariant correlation ratio:
\begin{equation}
    R^{O}_{\chi}  \equiv  1 - \frac{ \chi^O ( \bm{Q} + \Delta \bm{q} )  }{  \chi^O ( \bm{ Q } ) } 
\end{equation}
with using the largest eigenvalue $\chi^{O}(\bm{q})$ with $O$   referring  to SSC, QSH, and QSH$_z$, respectively. The ordering wave vector is $\bm{Q} = (0, 0)$, and $ | \Delta \bm{q}  | = \frac{4 \pi}{ \sqrt{3} L} $.     
We  note that since  the  superconducting order parameter  breaks  U(1)   global charge symmetry  and  the  QSH order parameter breaks  inversion  symmetry so that   the  conditions    for  inter-changing  the  limits   of  zero  temperature  and  infinite  size  are satisfied. 

\begin{figure}[htbp]
\centering
\includegraphics[width=0.41\textwidth]{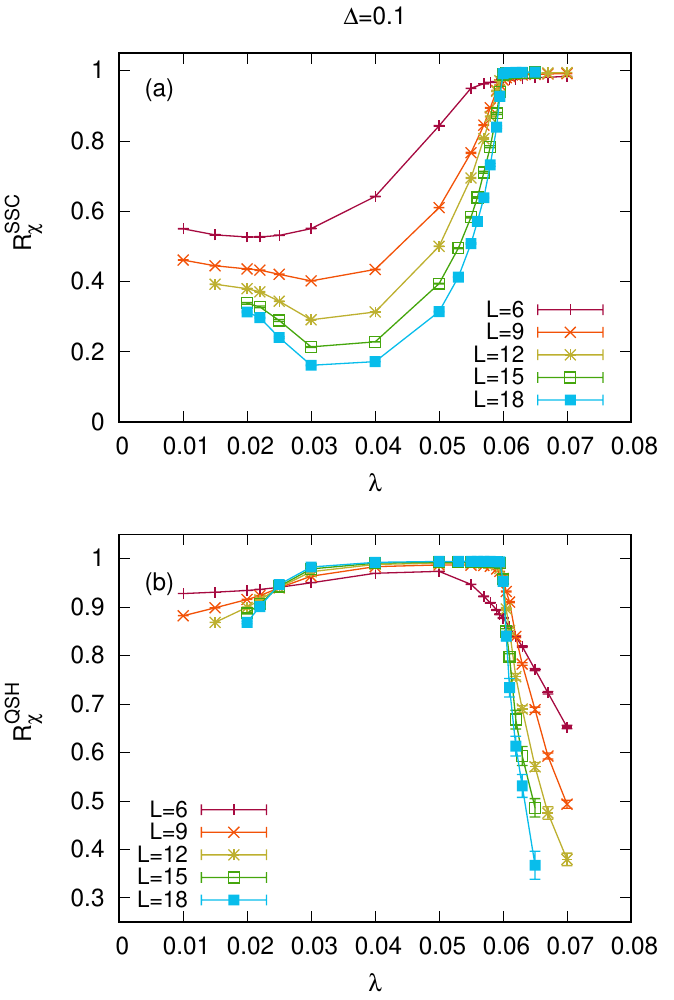}
\caption{\label{fig:Ratiotau_0.1}
Time displaced correlation ratio $R^{SSC}_{\chi}$ (a) and $R^{QSH}_{\chi}$ (b) as a function of $\lambda$ for $\Delta=0.1$. 
}
\end{figure}

\begin{figure}[htbp]
\centering
\includegraphics[width=0.41\textwidth]{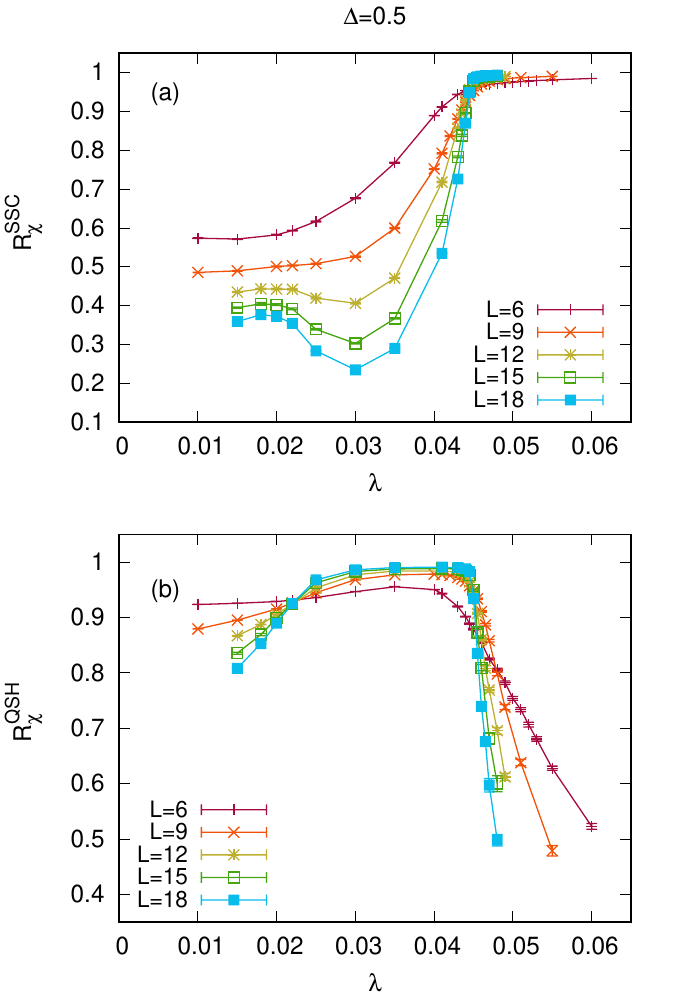}
\caption{\label{fig:Ratiotau_0.5}
Same as Fig.~\ref{fig:Ratiotau_0.1} for $\Delta=0.5$. 
}
\end{figure}

\begin{figure}[htbp]
\centering
\includegraphics[width=0.41\textwidth]{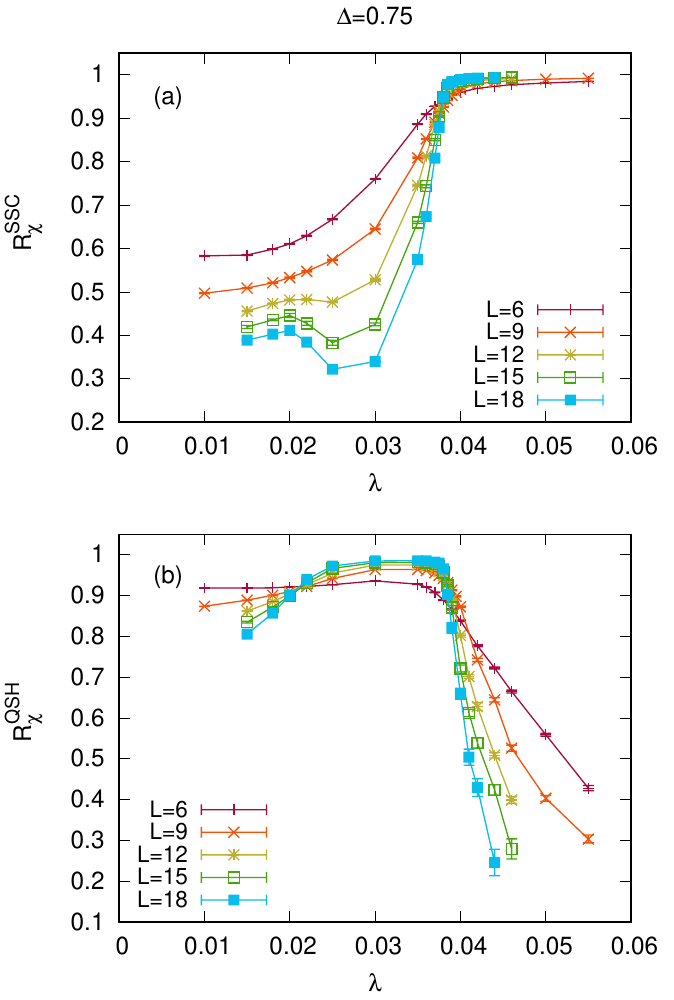}
\caption{\label{fig:Ratiotau_0.75}
Same as Fig.~\ref{fig:Ratiotau_0.1} for $\Delta=0.75$.  
}
\end{figure}

\begin{figure}[htbp]
\centering
\includegraphics[width=0.41\textwidth]{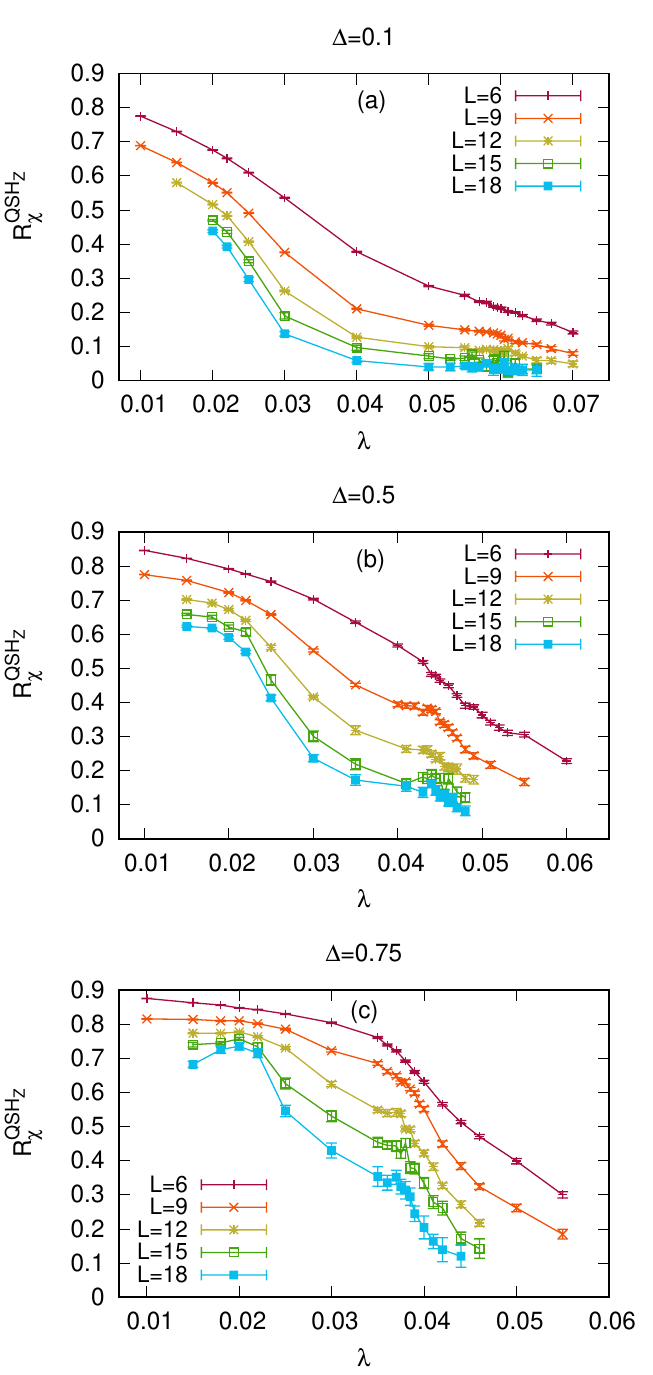}   
\caption{\label{fig:QSH_Z}
Time displaced correlation ratio $R^{QSHz}_{\chi}$ for $\Delta=0.1$ (a), $\Delta=0.5$ (b), and $\Delta=0.75$ (c), respectively.
}
\end{figure}

The finite size behavior of $R_{\chi}^{\text{QSH}}$
and $R_{\chi}^{ \text{SSC} }$ show no 
significant differences compared to the  corresponding  equal time correlations  considered in the main text.   
As shown in Figs.~\ref{fig:Ratiotau_0.1},  
\ref{fig:Ratiotau_0.5} and \ref{fig:Ratiotau_0.75}, the crossing points 
are consistent with the estimation of critical points from the main text.  
On the other hand, the scaling of $R_{\chi}^{\text{QSH}_Z} $ remains   
ambiguous: at the  QSH-SSC transition point, this quantity decays toward zero 
for strong anisotropy ($\Delta=0.1$).  In the case of 
$\Delta=0.5$ and $0.75$ it could converge to a finite constant in 
thermodynamic limit;  upon increasing the system sizes, its decreasing tendency is similar to the one from
equal time correlation ratio in the main text.

\clearpage
\section{Spectrum}
\label{App:Spectrum}

The  spectral  function for a  given  operator at  zero temperature reads:
\begin{equation}\label{Eq:spectrum}
    A(\bm{k},\omega) = \pi \sum_n |\langle n|\hat{O}|0\rangle|^2\delta(E_n-E_0-\omega).
\end{equation}
Here, $|0\rangle$ in Eq.~\ref{Eq:spectrum} denotes the ground state, and $ |n \rangle $ denotes  the eigenstates of the Hamiltonian with energy $E_n$.
Given the imaginary-time Monte Carlo data, the spectrum is  obtained using the  stochastic  analytical continuation   approach \cite{Beach044a}  to
solve  for  $A(\bm{k},\omega)$, given $ \langle 0|\hat{O}^\dagger(\tau) \hat{O}(0)|0 \rangle$: 
\begin{equation}
    \langle 0|\hat{O}^\dagger(\tau) \hat{O}(0)|0 \rangle = \int d\omega e^{-\tau \omega}A(\bm{k},\omega). 
\end{equation}
Here,  $\hat{O}$  represents the momentum space 
operators defined in  the  main text: ${\hat{J}}^{XY}$, $\hat{\eta}$, $\hat{J}^{Z}$ and $\hat{c}$. 
$A(\bm{k},\omega)$ is the corresponding spectral function, denoted as
$A^{\text{QSH}}$, $A^{ \text{SSC}}$, $A^{ \text{QSHz} }$ and $A^{sp}$,   respectively.

\begin{figure*}[htbp]
\centering
\includegraphics[width=1\textwidth]{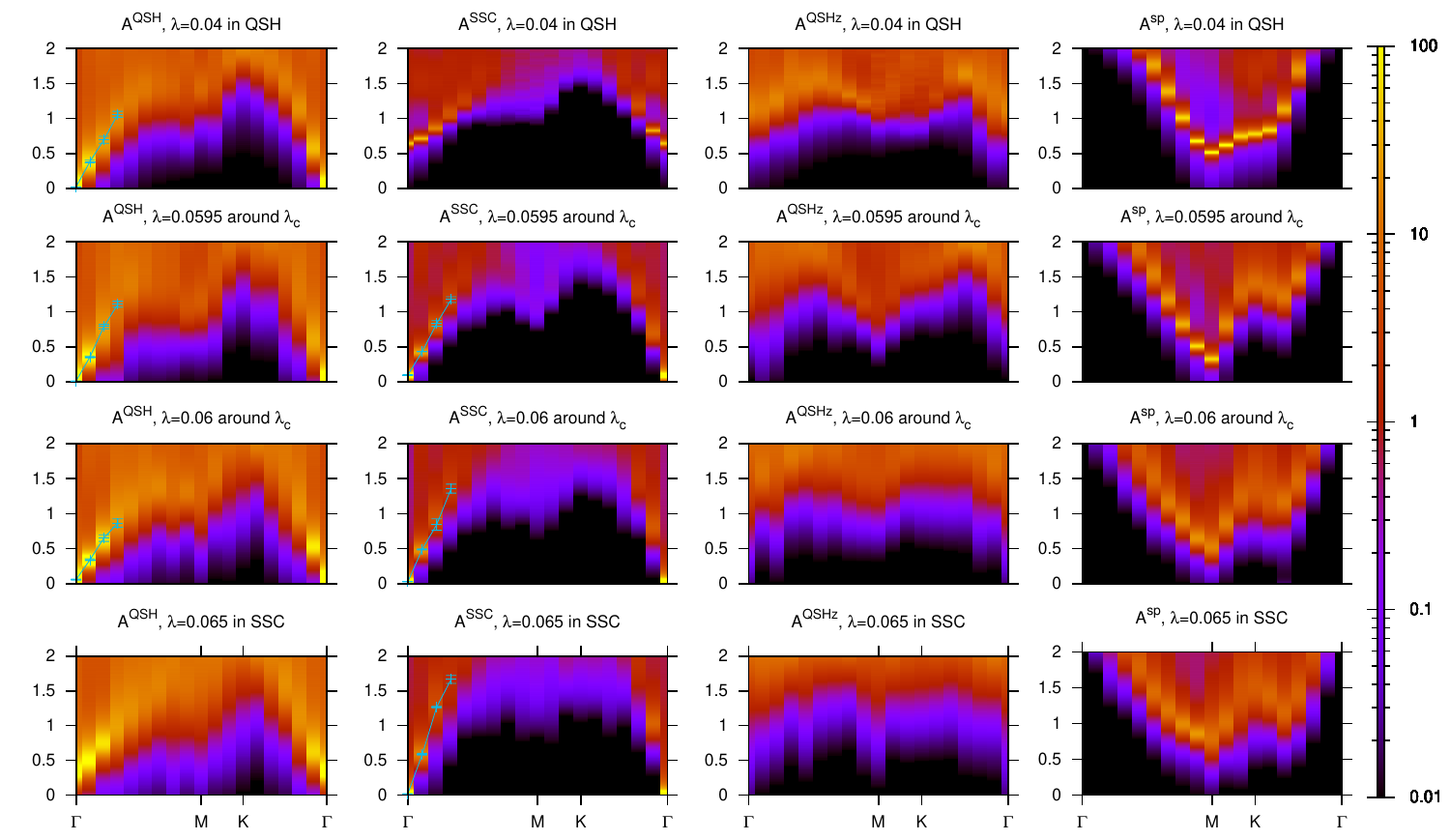}
\caption{\label{fig:dy0.1}
QSH, SSC, QSHz, and single particle spectrum inside two (QSH, SSC) phases and near the critical point, for $\Delta=0.1$. Blue lines 
are the momentum dependence of the extrapolated excitation gap of 
$ \hat{J} $ and $ \hat{\eta}^+(\hat{\eta}^{-} ) $  
operators. 
We took $L=18$. 
}
\end{figure*}

\begin{figure*}[htbp]
\centering
\includegraphics[width=1\textwidth]{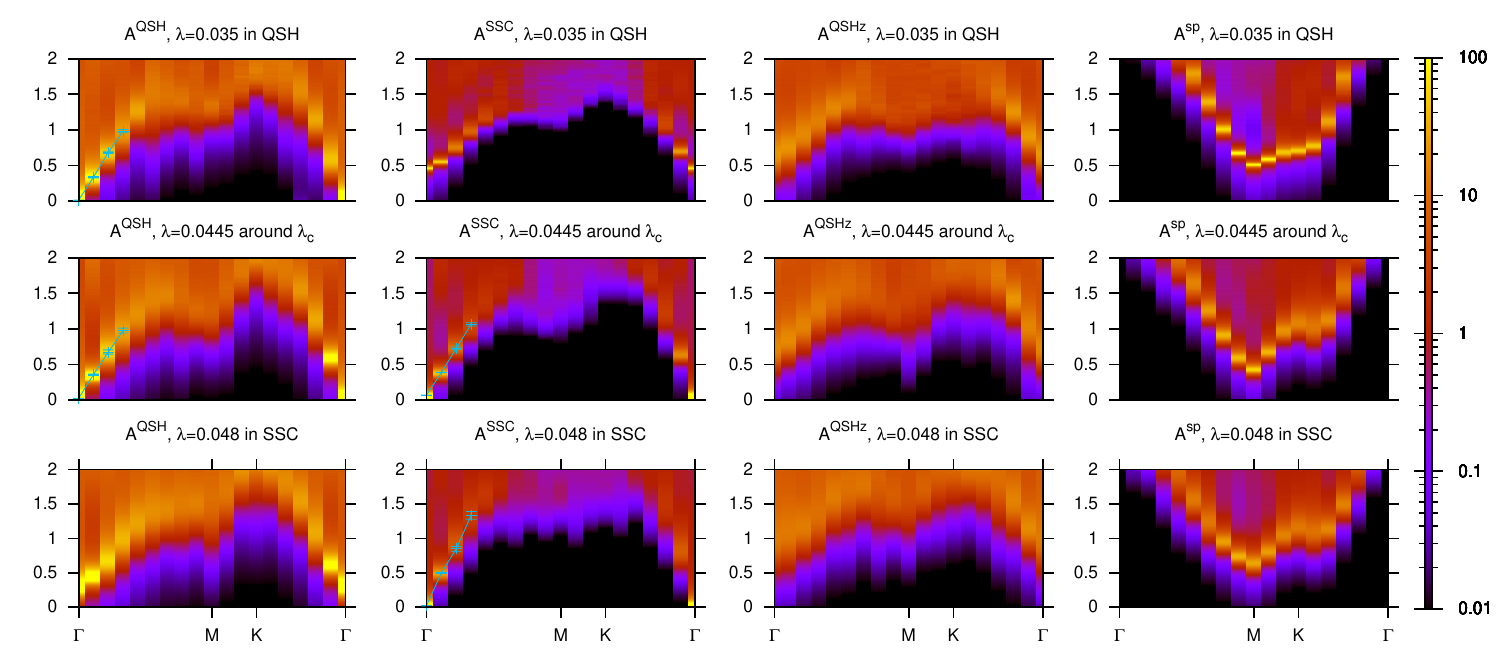}
\caption{\label{fig:dy0.5}
Same as Fig.~\ref{fig:dy0.1} for $\Delta=0.5$. 
}
\end{figure*}

\begin{figure*}[htbp]
\centering
\includegraphics[width=1\textwidth]{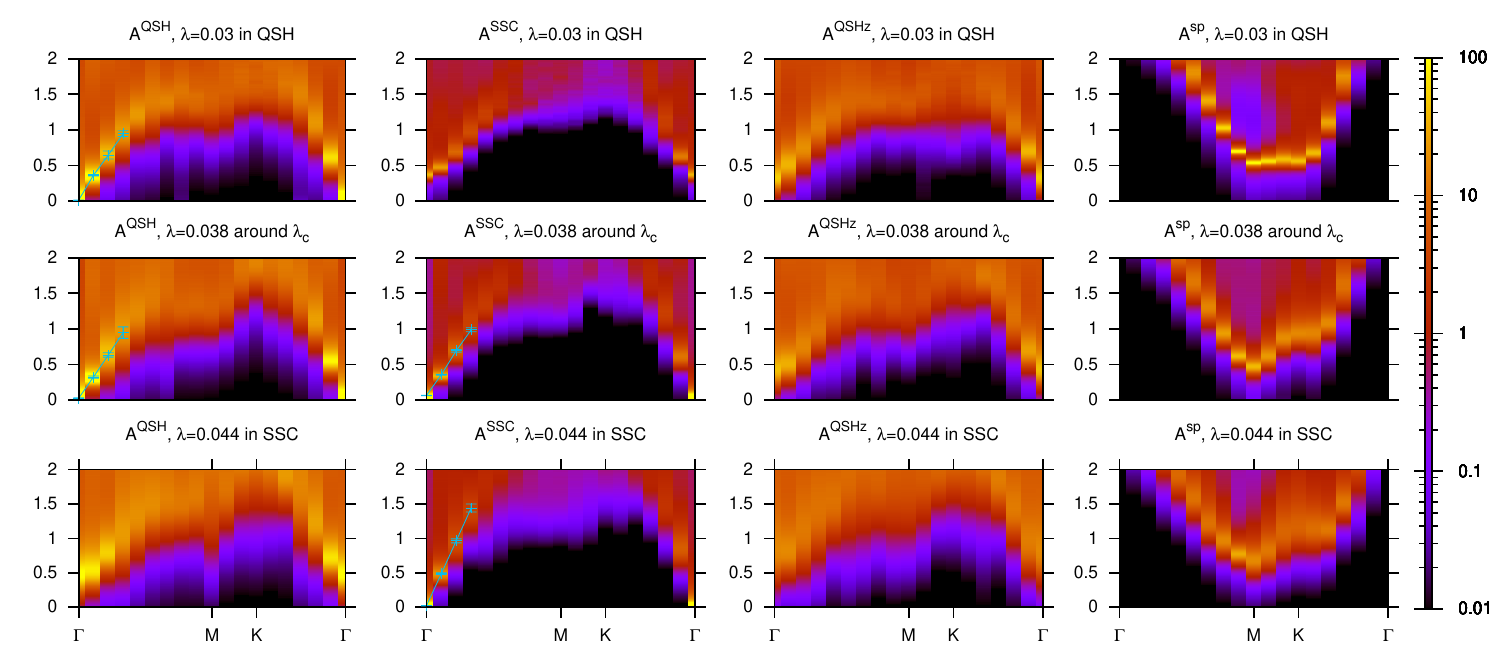}
\caption{\label{fig:dy0.75}
Same as Fig.~\ref{fig:dy0.1} for $\Delta=0.75$. 
}
\end{figure*}

As expected, the single-particle spectrum $A^{sp}$ is clearly gapped    
in both the  QSH and SSC ordered states, as well as   across  the  
QSH-SSC transition points.  As  one can observe  from  Figs.~\ref{fig:dy0.1}, \ref{fig:dy0.5}  
and \ref{fig:dy0.75}, $A^{sp}$    shows  no fundamental  
 differences among three   considered  values of $\Delta$.

Deep inside the ordered QSH and SSC  phases,   and irrespective of 
the  anisotropy parameter, $\Delta$,   
the  order  parameter excitations  exhibit  Goldstone  modes    stemming  
from  global  U(1)  symmetry  breaking. 
In  particular, in  Fig.~\ref{fig:dy0.1},    
for  the case  $\Delta=0.1$,   
a  linear mode is  observed     
for QSH operator at $ \lambda = 0.04 $ and for SSC operator at 
$\lambda = 0.065$. The  same behavior is visible   at $\Delta=0.5$ and $0.75$ in 
Fig.~\ref{fig:dy0.5} and \ref{fig:dy0.75}.

On the other hand,   
the excitation of both QSH and SSC order parameters at the critical point shows a linear dispersion 
relation. 
Near  the $\Gamma$ point,   the  Goldstone  mode  is   expected to  give  rise  to a branch  cut  reflecting the  
anomalous  dimension:
\begin{equation}
\begin{aligned}
 A_{\text{QSH}}(\bm{k},\omega) & \propto
 (v^2 |\bm{k}|^2 -\omega^2 )^{ 1 -\frac{\eta_{\text{QSH}}}{2}}  \\ 
 A_{\text{SSC}}(\bm{k},\omega) & \propto 
 (v^2 |\bm{k}|^2 -\omega^2 )^{ 1 -\frac{\eta_{\text{SSC}}}{2}}.
\end{aligned}
\end{equation} 
Although  the anomalous dimension of the two order parameters can in general be different,  
the velocity $v$ is uniquely defined at a Lorentz invariant critical point.    
We mark the velocity of these two excitations  
in Fig.~\ref{fig:dy0.1}, \ref{fig:dy0.5}  and \ref{fig:dy0.75}.

The spectrum of  the $Z$ component of the spin current operator ($ \hat{J}^{ Z } $) is 
controversial at  the  transition point.  
$ A^{ \text{QSH}_Z } ( \bm{k},\omega) $ shows a clear gap around  the 
$ \Gamma $ point in the cases of $ \Delta=0.1 $ and $0.5$. On the other 
hand, we   observe  that the gap decreases upon  
reducing the anisotropy.  As shown in Figs.~\ref{fig:dy0.75} and \ref{fig:SW_QCP}   for $\Delta=0.75$, the value of gap   at 
 the  $\Gamma$ point is  comparable  to  the  finite  size  gap of  the superconducting  fluctuations. 
A  consistent  picture  in  terms of easy-plane  DQCP  with  deconfined  spinons at    criticality  requires  
that the excitation gap of $ \hat{J} ^Z $  to scale to zero in the thermodynamic limit.

\section{Local detection of $Z_2$ topology using $\pi$ flux insertion}    
\label{App:Flux}

To detect the topology of our QSH insulator,  we employ the   
magnetic flux insertion approach~\cite{Qi08}  
that has successfully been tested in Ref.~\onlinecite{Assaad12}. 
When $\pi$ fluxes  are pumped  locally into the QSH insulator,   
mid-gap states carrying nontrivial spin    
quantum numbers are  exponentially localized around the flux insertion points. 
This approach  directly  probes  the  $Z_2$  topological invariant,  and  we refer  the  reader  to   Ref.~\onlinecite{Assaad12}  for 
a  detailed  discussion. 

Consider the following kinetic Hamiltonian: 
\begin{equation}\label{Eq:Ham_T_flux}
\begin{aligned}
 \hat{H}_t   = - t  \sum_{ \langle \bm{i}, \bm {j} \rangle } (\hat{\ve{c}}^{\dagger}_{\bm{i} } \hat{\ve{c}}^{\phantom\dagger}_{\bm{j}}  e^{ i A_{\bm{i} \bm{j}} } + H.c.).
\end{aligned}
\end{equation} 
and the interaction term:
\begin{equation}\label{Eq:Ham_V_flux}
\begin{aligned}
 \hat{H}_{\lambda}  = & -\lambda \sum_{\varhexagon}
   \left[  \left( \sum_{\langle \langle \bm{i} \bm{j} \rangle \rangle  \in \varhexagon }   \hat{J}^x_{\bm{i},\bm{j}}  e^{ i A_{\bm{i} \bm{j}} } \right)^2
   +    \left( \sum_{\langle \langle \bm{i} \bm{j} \rangle \rangle  \in \varhexagon }   \hat{J}^y_{\bm{i},\bm{j}}   e^{ i A_{\bm{i} \bm{j}} } \right)^2     \right.      \\
 &  \phantom{=\;\;} \left.
 +  \Delta  \left( \sum_{\langle \langle \bm{i} \bm{j} \rangle \rangle  \in \varhexagon }   \hat{J}^z_{\bm{i},\bm{j}}  e^{ i A_{\bm{i} \bm{j}} } \right)^2     \right]
\end{aligned}
\end{equation}
where $ \bm{A}_{ \bm{i} \bm{j} } $ is the vector potential  that accounts  for  the pair of $\pi$-fluxes.    
To practically insert a $\pi$-flux into our system, we consider an arbitrary string connecting the centers of the two hexagons.
Each  time  an  electron crosses    this 
string, it  acquires an  $e^{i \pi}$  phase  factor.   

\begin{figure}
    \centering
    \includegraphics[width=0.54\textwidth]{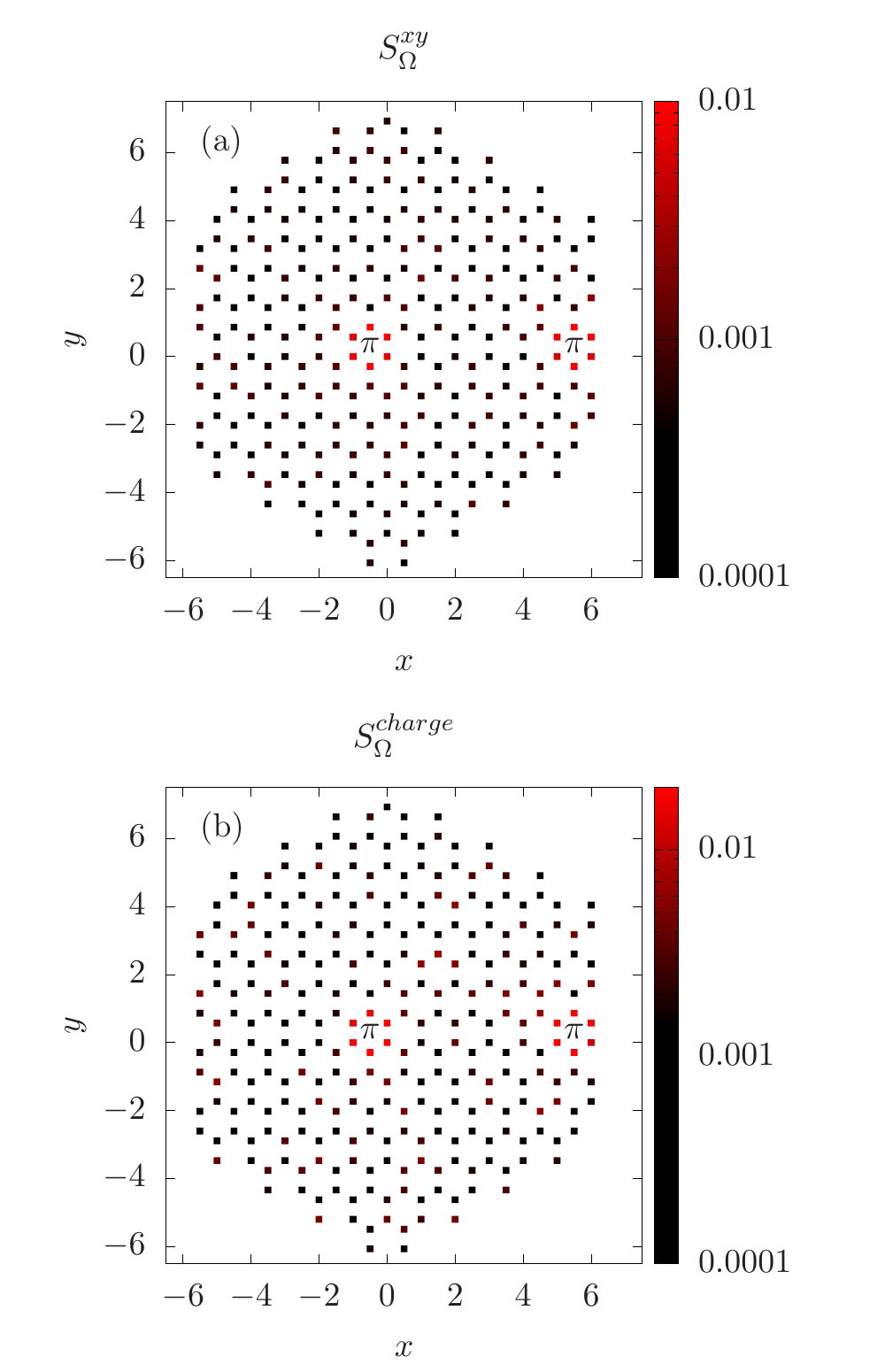} 
    \caption{ $ S^{x y}_{\Omega} (\bm{i}) $ (a)  and  $ S^{\text{charge}}_{\Omega} (\bm{i}) $ (b) 
    distributions on a  
    $L=12$ honeycomb lattice. 
    The simulation is performed deep inside the QSH state ($\Delta=0.1,  \lambda=0.045$).
    Here, $\beta=24$.  }
    \label{fig:S_O}
\end{figure}

Due to the easy-plane anisotropy,     
 the dynamical generation of the QSH insulator is associated with a long-range 
order of  spin currents in the $U(1)$ plane. 
Thus, the mid-gap objects localized around the $\pi$ fluxes are  
Kramers pairs of 
spin `up' and `down' states rotating in the $x-y$ plane, 
as well as doublets of charge fluxons.  
The  presence of localized spin and charge fluxons can be captured by the low-energy 
spectral weight of $ c^{\dagger}_{\bm{i}}  \sigma^x c^{\phantom\dagger}_{\bm{i}} $ ($ c^{\dagger}_{\bm{i}}  \sigma^y c^{\phantom\dagger}_{\bm{i}} $) and $c^{\dagger}_{\bm{i}}c^{\phantom\dagger}_{\bm{i}}$ operator:  
\begin{equation}
\begin{aligned}
 & S^{x y}_{\Omega} (\bm{i})  \equiv  \int_0^{\Omega}  d \omega   S^{x y} ( \bm{i}, \omega )  
     \\   
 & S^{\text{charge}}_{\Omega} (\bm{i})  \equiv  \int_0^{\Omega}  d \omega   S^{\text{charge}} ( \bm{i}, \omega )  
  \  \  \quad  \Omega = 0.25   
\end{aligned} 
\end{equation}
where 
\begin{equation}
\begin{aligned}
& S^{x y} ( \bm{i}, \omega ) = \pi \sum_n  | \langle n | c^{\dagger}_{\bm{i}} 
  \sigma^x c^{\phantom\dagger}_{\bm{i}} | 0 \rangle |^2  \delta ( \omega - E_n - E_0  )    \\ 
& S^{\text{charge}} ( \bm{i}, \omega ) = \pi \sum_n  | \langle n | c^{\dagger}_{\bm{i}} 
   c^{\phantom\dagger}_{\bm{i}} | 0 \rangle |^2  \delta ( \omega - E_n - E_0  )   
\end{aligned}    
\end{equation}
where $ |n\rangle$  represents  an  energy  eigenstate   with energy $E_n$.  The energy window of 
$\Omega=0.25$ is well below twice single particle gap ($\Delta \approx 1.0$).

The  enhanced  spectral  weight  in $S^{x y}_{\Omega} (\bm{i})$ and $S^{\text{charge}}_{\Omega} (\bm{i})$ around $\pi$ fluxes, 
as depicted in  
Fig.~\ref{fig:S_O}, clearly  demonstrates  the existence of spin and charge fluxons.   
This numerically proves that the insulating state that we  observe  at   intermediate  values of   
$\lambda$ is $Z_2$ nontrivial. 

It is worth noting that  our  model does not exhibit quantized spin Hall conductivity due to the absence of $U(1)$ spin conservation at low energies.  
Therefore, the  topological invariant characterizing our system is the $Z_2$ index, rather than the  so-called  spin  Chern number.

\section{Absence of monopoles}
\label{monoploes.sec}

The aim of this section is to show the absence of monopoles in our system. A  
`monopole'   corresponds  to a singularity of the $U(1)$ gauge field in CP$^1$ representation.
It  couples to  the physical electric-magnetic vector potential such that  
 a $U(1)$ electric-magnetic gauge invariance is broken  in the presence 
 of `monopoles'~\cite{Grover08}.   Therefore, the statement of `monopole-free' is a natural consequence of the 
 exact  
 $U(1)$ charge conservation in our physical system.

A continuum description of our system in terms of fluctuating QSH order parameter in $S^2$ space is: 
\begin{equation}
\begin{aligned}
  S = &\int d^2 \bm{x} d \tau  \psi^{\dagger} [  \gamma_0 \gamma_{\mu}  (  \partial_{\mu}  + i e A_{\mu } ) \\ 
  + &
   i m  \ve{n} \cdot \gamma_0 \gamma_3 \gamma_5  \vec{\sigma}    ]   \psi     
   + S_{ \text{anisotropy}}  + ... 
\end{aligned}
\end{equation} 
where $ \psi( \bm{x}, \tau )$ is the Grassmann variable in the space of
$ \mathds{R}^2 \otimes \mathds{C}^2_{valley} \otimes \mathds{C}^2_{orbital} \otimes  \mathds{C}^2_{spin}  $.  
$ \ve{n}$ is the three-component order parameter of  the QSH state and  
$S_{ \text{anisotropy}}$  describes  the anisotropic term which breaks SU(2) spin rotational 
symmetry down to U(1)$\times  {\mathbb Z}_2 $. 
 $A_{\mu}(\bm{x}, \tau) $  is the physical electric-magnetic field. 
 Note that our lattice model doesn't break particle number conservation,    
 such that   
  local $U(1)$ gauge symmetry  is  satisfied:  
\begin{equation}\label{Eq:U1_local}
\begin{aligned}
 & \psi( \bm{x}, \tau ) \rightarrow e^{ i e \theta(\bm{x}, \tau) }  \psi( \bm{x}, \tau),   \\ 
 & A_{\mu}(\bm{x}, \tau) \rightarrow A_{\mu}(\bm{x},\tau) -  \partial_{\mu} \theta( \bm{x}, \tau ). 
\end{aligned}
\end{equation} 
    
Since  $\ve{n}$  is  normalized  to  unity,  the  single particle  gap does not close, and we can integrate out the fermions  to obtain  the  effective action: 
\begin{equation}
    S = S_g +  S_{c}
\end{equation}
where  
\begin{equation}
  S_g = \int d^2 \bm{x} d \tau 
  \frac{1}{m} [ (  \partial_{ \mu } \ve{n} ) \cdot ( \partial_{\mu } \ve{n} )   ] 
\end{equation}
and the lowest-order electric-magnetic response is  
\begin{equation}
\begin{aligned} 
 S_c = i \int  d^2 \bm{x} \int d\tau  A^{ \mu } (\bm{x},\tau) J^{\mu} (\bm{x}, \tau )  
\end{aligned}    
\end{equation}
where 
\begin{equation}
\label{Jmu}
  J^{\mu} (\bm{x}, \tau ) = 2 e  \frac{1}{ 8 \pi }   \epsilon^{ \mu \alpha \delta } 
  \ve{n} \cdot ( \partial_{\alpha} \ve{n}  \times \partial_{\delta} \ve{n} )
\end{equation}

Now reformulate the action into a gauge redundant (CP$^1$) representation:  
\begin{equation}
    \vec{N}  \equiv  z^{\dagger}  \vec{\sigma}  z   \quad \quad \quad z^\dagger z = 1
\end{equation} 
such that $S_g$ is described by $\frac{1}{g} |  ( \partial_\mu - i a_{\mu} )z |^2 $.    
A local $U(1)$ gauge redundancy comes naturally from the $CP^1$ reformulation:   
\begin{equation}
     z \longrightarrow e^{i \chi } z,  \  \  \  \  a_{\mu} \longrightarrow  a_{\mu} + \partial_{\mu} \chi  
\end{equation}

Crucially  the following identity holds: 
\begin{equation}\label{N_to_a}
\begin{aligned}
         \frac{1 }{ 8 \pi } \epsilon^{ \mu \nu \lambda }  \ve{n} \cdot ( \partial_{\nu} \ve{n}  \times  \partial_{\lambda}   \ve{n} )         
 =    \frac{ 1 }{ 4 \pi }  \epsilon^{ \mu \nu \lambda }  \partial_\nu a_{\lambda}.  
\end{aligned}    
\end{equation} 
This follows  from the the saddle point   results  \cite{Fradkin13}:
\begin{equation}
   a_{\mu} =   i z^{\dagger} \partial_{\mu} z. 
\end{equation} 
To describe   skyrmions   one  has   to allow    for compactness of  $ a_{ \mu } $ :    
\begin{equation}
    \oint \ve{a} \cdot d \ve{l}  =   n 2\pi   \, \,  \text{  with }   \, \,  n \in \mathbb{Z}
\end{equation}
 for  a  closed  loop. 
Using   Stokes    theorem,  integration at  a  given   time slice   hence gives:  
$  \int d^2 \bm{x} \epsilon^{0 \mu \nu } ( \partial_\mu  a_\nu  -  \partial_\nu  a_\mu  ) =  4 \pi  n $   and  counts 
the number  of  `skyrmions' in  the  time slice.  
A magnetic   `monopole'     corresponds  to field  configurations,   $a_{\mu}$,    
that  have  a  non-vanishing   `magnetic flux',  $  \epsilon^{\delta \mu \nu } \partial_\mu  a_\nu $, 
 through a closed surface in space-time.

With  Eq.~\ref{N_to_a}    and  Eq.~\ref{Jmu},  the Chern-Simons part of action reads
\begin{equation}
  S_c =   \int d^2 \bm{x} d \tau  
    \frac{ i e }{ 2 \pi } \epsilon^{ \mu \nu \kappa }  A_{\mu} \partial_{\nu} a_{\kappa}     
\end{equation}

To  show   the  absence of  monopoles,   we consider  a   uniform gauge transformation inside a  sphere:    
\begin{equation}
    A_{ \mu } \longrightarrow  A_{ \mu } +  \partial_{\mu}  \Theta    
 \end{equation}
 with
 \begin{equation}
    \Theta( \tau, x, y )    =   
\left\{
\begin{array}{ll}
 \Theta_0    &   \text{   for }    r <R   \\ 
              0   &   \text{   for  }    r \geq R    
\end{array}                                                
\right. .
\end{equation}
In the  above  $r   =  \sqrt{\tau^2 + x^2 +   y^2} $.    
As  a  consequence,   $ \partial_{\mu}  \Theta  $ does  not   vanish  only  on sphere  of  radius  R,  $S_R$.  In particular: 
\begin{equation}
\begin{aligned}
      \delta S = &   \frac{i e }{ 2\pi } \int d^{3} \ve{x}   \epsilon^{ \mu \nu \kappa }  \partial_{\mu} \Theta  \,  \partial_{\nu}  \, a_{\kappa} 
                    \\ 
= & \frac{i  e}{ 2\pi }   \Theta_0 \int_{S_R}     d  \ve{s}_{\mu}       \epsilon^{ \mu \nu \kappa }   \partial_{\nu}  \, a_{\kappa}   \\
\equiv  &   \frac{i  e}{ 2\pi }   \Theta_0   Q. 
\end{aligned}
\end{equation}
In the   above,   $d \ve{s}$   defines  a  surface  element of  $S_R$  and  $Q$  the   number of  monopoles   within 
$S_R$.     If  $Q  \neq 0$  then U(1)  local  gauge   invariance    (see Eq.~\ref{Eq:U1_local} )  is  not  satisfied.  Hence,
charge  conservation,  or  equivalently  U(1)  local  gauge   invariance,    requires   $Q=0$   and  the absence of  monopoles.

It is important to emphasize that, based on spin coherent path integral calculation by  
Haldane,~\cite{Haldane88}  the monopole configurations of  anti-ferromagnetically coupled spin $1/2$ system on the  
 square lattice also  
carry a nontrivial phase factor upon a $U(1)$ gauge transformation.  
However,  in this case, instead of  the physical electromagnetic field, this $U(1)$ transformation corresponds 
to the lattice  rotation.  This symmetry  
is broken  upon lattice regularization, and   
only a $Z_4$ subgroup of it is  conserved.    
Under $\pi/2$  lattice rotation,  a single monopole contributes  by 
\begin{equation}
  \delta S = i \frac{\pi}{2} 
\end{equation} 
to the action. Hence,  quadruple  monopoles   configurations  are  allowed.

\end{document}